\documentclass[acmlarge, screen, pdfusetitle, pdfdisplaydoctitle]{acmart}
\AtBeginDocument{%
  \providecommand\BibTeX{{%
    \normalfont B\kern-0.5em{\scshape i\kern-0.25em b}\kern-0.8em\TeX}}}

\usepackage{enumerate}
\usepackage{graphicx}
\usepackage{subcaption}
\usepackage{float}
\usepackage{multirow}
\usepackage{array}
\usepackage{makecell}
\usepackage{gensymb}
\usepackage{enumitem}
\usepackage{caption}
\definecolor{lightgray}{gray}{0.95}  

\setcopyright{cc}
\setcctype{by}
\acmJournal{IMWUT}
\acmYear{2026} \acmVolume{10} \acmNumber{4} \acmArticle{262}
\acmMonth{12} \acmDOI{10.1145/3857922}

\newcommand{\system}{NavSight}

\definecolor{brownishred}{RGB}{178, 34, 34}

\DeclareRobustCommand{\camchange}[1]{%
  #1%
}

\begin{document}

\title[NavSight]{NavSight in the Wild: Understanding Real-World Use of a Mobile Augmented Reality Application for People with Low Vision in Outdoor Navigation}

\author{Yuheng Wu}
\correspondingauthor
\orcid{0009-0005-1828-400X}
\affiliation{%
  \department{Department of Computer Sciences}
  \institution{University of Wisconsin-Madison}
  \city{Madison}
  \state{WI}
  \country{USA}
}
\email{yuheng.wu@wisc.edu}

\author{Kexin Zhang}
\orcid{0009-0009-4078-8780}
\affiliation{%
  \department{Department of Computer Sciences}
  \institution{University of Wisconsin-Madison}
  \city{Madison}
  \state{WI}
  \country{USA}
}
\email{kzhang284@wisc.edu}

\author{Ben Kosa}
\orcid{0009-0006-5836-4925}
\affiliation{%
  \department{Department of Computer Sciences}
  \institution{University of Wisconsin-Madison}
  \city{Madison}
  \state{WI}
  \country{USA}
}
\email{bkosa@cs.wisc.edu}

\author{Ru Wang}
\orcid{0000-0003-3907-2370}
\affiliation{%
  \department{Department of Computer Sciences}
  \institution{University of Wisconsin-Madison}
  \city{Madison}
  \state{WI}
  \country{USA}
}
\email{ru.wang@wisc.edu}

\author{Sanbrita Mondal}
\orcid{0000-0003-4454-8978}
\affiliation{%
  \department{School of Medicine and Public Health}
  \institution{University of Wisconsin-Madison}
  \city{Madison}
  \state{WI}
  \country{USA}
}
\email{smondal4@wisc.edu}

\author{Yuhang Zhao}
\orcid{0000-0003-3686-695X}
\affiliation{%
  \department{Department of Computer Sciences}
  \institution{University of Wisconsin-Madison}
  \city{Madison}
  \state{WI}
  \country{USA}
}
\email{yuhang.zhao@cs.wisc.edu}

\renewcommand{\shortauthors}{Wu et al.}

\begin{abstract}
The ability to navigate outdoors safely and independently is crucial yet challenging for people with low vision (PLV). While various augmented reality (AR) systems for low vision have been designed and evaluated in ideal lab environments, no research has investigated their real-world feasibility and challenges. We present \textit{\system{}}, a mobile AR application that assists PLV in outdoor navigation by recognizing important outdoor objects (e.g., curb, vehicle) and rendering real-time visual augmentations. 
Through a seven-day diary study with 12 PLV in real-world settings, we characterize the impact of \system{} on scene perception, users' configuration strategies on what objects to augment and how to augment them across scenarios, how users made sense of and responded to recognition errors, and the social acceptability of using \system{} in public.
We further identify environmental factors affecting recognition, such as weather conditions, lighting and shadows, and nonstandard road markings and textures, as well as usability issues in daily use. We discuss these real-world challenges and derive design implications for future AI-powered assistive AR systems for outdoor use.
\end{abstract}




\maketitle

\section{Introduction}
\label{sec:introduction}
The ability to navigate outdoors safely and independently is fundamental for everyday activities, such as hiking in a park or traveling to grocery stores. However, it remains a major challenge for people with low vision (PLV) \cite{zeng2015survey,szpiro2016shopping, muller2022traveling}, who experience vision loss but still have remaining vision to use \cite{nihVisionNational, szpiro2016shopping}.

Various augmented reality (AR) systems have been designed to support safe and efficient navigation for PLV through different visual augmentations \cite{zhao2020indoor, fox2023using, maman2025enhancing, Htike2021Augmented, hicks2013depth, angelopoulos2019enhanced, Sadeghzadeh2024ARVA, chen2025visimark, sayed2020mobility}. Zhao et al. \cite{zhao2020indoor} render visual guidance cues (e.g., path visualizations and signs indicating route distance and next actions) to facilitate wayfinding.
Obstacle-aware systems \cite{fox2023using, maman2025enhancing, Htike2021Augmented} augment obstacles along the route (e.g., outlining their contours) to support obstacle avoidance. Depth-based systems \cite{hicks2013depth, angelopoulos2019enhanced, van2015improving} recolor the scene based on depth to enhance PLV's depth perception during navigation. Finally, Sadeghzadeh et al. \cite{Sadeghzadeh2024ARVA} and Sayed et al. \cite{sayed2020mobility} remap the visual scene to compensate for field-of-view loss, enabling a broader view of the surroundings for obstacle avoidance and sign reading.
However, all of these systems were evaluated in ideal, indoor lab environments for short sessions with predefined tasks, such as walking down a predefined path (e.g., a corridor) \cite{maman2025enhancing, fox2023using, Sadeghzadeh2024ARVA, zhao2020indoor, angelopoulos2019enhanced, hicks2013depth, sayed2020mobility} or exploring a pre-set room with certain obstacles \cite{Htike2021Augmented}.
It remains unclear how these systems could generalize to \textit{uncontrolled real-world outdoor} navigation. 

Visual augmentations for PLV in AR face unique challenges in real-world outdoor environments. For example, the visibility of AR elements may degrade in sunlight \cite{xiong2021augmented, erickson2022review, szpiro2016shopping}. PLV may need additional time for vision adjustment when transitioning between indoor and outdoor environments \cite{szpiro2016shopping}, which may lead to different AR augmentation needs during that transition. 
Moreover, PLV's visual preferences and needs may also change across scenarios (e.g., prioritizing obstacle avoidance and street crossing during navigation vs. locating and reading signs when approaching the destination) \cite{szpiro2016shopping, gamage2023blind, starke2020everyday} and over time \cite{zolyomi2017technology, phillips1993predictors}. However, these important perspectives that affect AR technology usage and adoption are usually not captured by short-term, indoor lab studies. 
As a result, the ecological validity of these AR aids remains unclear---it is unknown whether they can remain helpful across diverse, uncontrolled everyday outdoor environments with varying scene complexity, dynamics, and user needs. 


To fill this gap and facilitate long-term adoption in the real world, we seek to investigate the ecological validity of AR navigation support for PLV via a diary study, revealing their evolving usage patterns, augmentation preferences, and unique challenges when using such technology in various real-world outdoor environments over time. To enable such a study, we developed \textit{\system{}}, a mobile AR application that recognizes and augments key outdoor objects (e.g., sidewalk, curb, pedestrian) in real time to support navigation \camchange{as a complement to PLV's existing mobility aids (e.g., white canes)}. The application provides six AR augmentations inspired by prior research \cite{zhao2016cuesee, zhao2015foresee, lee2024cookar, fox2023using, maman2025enhancing, gopalakrishnan2025comparison}. With \system{}, users with low vision have the flexibility to select what objects to augment and how to augment them, allowing us to capture their augmentation preferences and usage patterns in different scenarios. We chose mobile AR over wearable displays since smartphones are widely available \cite{pundlik2023impact}, socially acceptable \cite{goodman2019social, denning2014situ}, and possess sufficient computational power for real-time on-device processing to ensure responsiveness \cite{lee2019device}.

By deploying \system{}, we conducted a diary study with 12 participants with low vision, who flexibly used and customized the application in all kinds of outdoor environments over a week and filled out daily surveys to reflect on their experiences.   
Across the study, participants reported several ways \system{} supported their outdoor navigation, including simplifying outdoor scenes into walkable and non-walkable regions, helping them notice tripping hazards and approaching objects, and extending their visual reach. 
Some participants even used it in other outdoor activities beyond navigation, such as checking queue size or watching a soccer game. 
Participants continuously configured \system{}, selecting objects to augment based on what they expected to encounter and what they were doing, grouping objects by risk level, and adjusting augmentation selections to minimize occlusion and optimize visibility against the scenes.
However, due to different types of recognition errors and other usage challenges, participants could not fully rely on \system{} and thus had to divide their attention between the phone screen and their surroundings.

We identified real-world challenges that undermined participants' experiences. Augmentations became hard to see in bright sunlight due to screen glare and were obscured by reflections on wet surfaces after rain.
\system{}'s recognition of walkable paths was also affected by several real-world conditions, such as surface appearance changes caused by rain, inconsistent lighting and shadows, and nonstandard textures.  
Some participants formed their own mental models of recognition errors, attributing them to how they held the phone or how many objects they selected to augment, and those who could not form a coherent explanation trusted \system{} less.
In addition, while participants found \system{} less conspicuous than conventional aids (e.g., white canes), they faced the challenge of being misunderstood as filming.



With these findings, we contribute an empirical understanding of how PLV use AR augmentations in real-world outdoor navigation, along with design implications for future systems. This paper makes the following contributions: 
(1) \system{}, a deployable mobile AR application that recognizes 21 categories of important outdoor objects and renders real-time visual augmentations on smartphones. Users can flexibly configure what to augment and how by selecting objects to augment, assigning them to augmentation groups, and choosing augmentations for each group.
(2) An in-the-wild investigation of how PLV use AR visual augmentations for outdoor navigation. Through a seven-day diary study deploying \system{}, we characterize PLV's real-world usage scenarios, the ways augmentations reshaped their perception, the strategies they used to optimize object augmentation experiences across scenarios and over time, their interpretation of AI errors, and challenges and concerns about using \system{} in public.
(3) Design implications for future AI-powered AR systems as outdoor low-vision aids, grounded in participants' usage experiences and needs. 

\section{Related Work}
We motivate our work from three aspects: outdoor navigation tools for blind and low-vision people, AR technology for people with low vision, and field study of assistive technology for people with disabilities.


\subsection{Outdoor Navigation Tools for Blind and Low-Vision People}
Blind and low-vision (BLV) people frequently rely on white canes and guide dogs to detect obstacles, but these tools can only partially address outdoor navigation challenges. For example, white canes have a short range for ground-level obstacles and cannot identify obstacles above the waist \cite{dos2021electronic, dos2021systematic}. To overcome these barriers, researchers have proposed many assistive tools using different sensors to interpret users' environments \cite{Messaoudi2022Review, Real2019Navigation, Kuriakose2020Tools, el2021systematic}.
A large part of these tools focuses on obstacle avoidance. For example, prior work uses ultrasonic sensors \cite{patil2018design, shoval1998navbelt, bhatlawande2012ultrasonic, meshram2019astute, rahman2020obstacle},
RGB-D cameras \cite{xiao2015assistive, joseph2015being, joshi2020efficient, islam2020automated, yang2018unifying}, wearable RGB cameras \cite{duh2020v}, infrared sensors \cite{chaccour2015multisensor, chang2020design}, or radars \cite{kwiatkowski2017concept} to detect incoming obstacles and informs users via auditory or haptic cues.
GuideCane \cite{ulrich2001guidecane} further incorporated motors to automatically maneuver around obstacles. In addition to obstacle avoidance, some assistive systems also help with other navigation challenges, such as detecting water puddles and wet surfaces \cite{patil2018design, meshram2019astute, sahoo2019design, nandini2019smart}, avoiding potholes \cite{ulfa2023s, nandini2019smart, rao2016vision, chun2019ground}, wayfinding \cite{duh2020v}, or crossing intersections \cite{cheng2018intersection, tian2021dynamic, cheng2017crosswalk, cheng2018real}.

Smartphones further expanded the opportunities for assisting BLV people with outdoor navigation, as they are portable, convenient, equipped with a wide range of sensors for location, motion, and environmental data, and capable of high-performance computation \cite{Kuriakose2020Tools, kuriakose2020smartphone}. Leveraging these features, researchers have proposed smartphone-based applications to facilitate obstacle avoidance \cite{kuriakose2020smartphone, bai2017cloud, dutta2018divya, nawin2018navtu}, wayfinding \cite{velazquez2018outdoor, nawin2018navtu, chen2026navinote}, pedestrian signal detection \cite{yu2019street, ghilardi2018real, shangguan2014crossnavi}, and crosswalk detection \cite{yu2019street, murali2013smartphone} through audio and haptic feedback.

Unlike people who are totally blind, PLV rely heavily on their residual vision for daily activities \cite{szpiro2016shopping}. This reliance highlights the need for assistive technologies that enhance visual information rather than provide only non-visual feedback, leading to the development of augmented reality (AR) assistive technologies that visually augment PLV's residual vision for low vision assistance.

\subsection{AR Technology for People with Low Vision}

Recent years have seen the rapid development of AR technology as low-vision aids due to its ability to directly enhance visual information in various environments and tasks \cite{azenkot2017designing, zhao2017understanding}.
Early systems applied image processing techniques to the real-time camera feed of users' environments to enhance visual elements \cite{deemer2018low}, such as magnification \cite{zhao2019designing, stearns2018reading, zhao2015foresee}, edge enhancement \cite{hwang2014augmented, kwon2012contour, zhao2015foresee}, contrast enhancement \cite{zhao2015foresee}, and pixel remapping \cite{luo2006use, zhao2019computational, loshin1989programmable}.
More recent work augments only task-relevant objects to more effectively support daily activities \cite{zhao2016cuesee, lee2024cookar, zhao2019arstairs, huang2019augmented, lang2021pressing}.
For example, CueSee \cite{zhao2016cuesee} supported visual search by rendering visual cues (e.g., flashing contours) on the target objects, and CookAR \cite{lee2024cookar} facilitated kitchen tool interaction by augmenting tool affordances (i.e., graspable vs. hazardous parts).

For navigation, prior research has explored different augmentations to support obstacle avoidance and wayfinding. Depth-based systems recolored the scene by distance to convey depth \cite{hicks2013depth, angelopoulos2019enhanced, van2015improving}, and Sadeghzadeh et al. \cite{Sadeghzadeh2024ARVA} and Sayed et al. \cite{sayed2020mobility} remapped the scene to fit a broader view of the surroundings into a user's residual field of view.
Maman et al. \cite{maman2025enhancing} introduced a Wizard-of-Oz AR system that highlights obstacles with pink outlines for obstacle detection and avoidance. Similarly, Fox et al. \cite{fox2023using} utilized high-contrast outlines anchored to predetermined obstacles and head-attached arrows pointing to the obstacles to support avoidance. Further expanding the design space, Htike et al. \cite{Htike2021Augmented} explored eight augmentation prototypes, such as colored overlays, line overlays, and path visualizations, to facilitate obstacle avoidance and indoor wayfinding. Zhao et al. \cite{zhao2020indoor} explored visual and audio wayfinding guidance to improve navigation accuracy and lower cognitive load.
Jo et al. \cite{jo2023enhancing} designed a map guidance interface that displays routes and map overviews on users' peripheral vision to mitigate central vision loss during wayfinding.
However, all these systems were evaluated in ideal lab environments with predefined tasks.

Deploying visual enhancement AR systems for everyday use in uncontrolled outdoor scenarios involves unique challenges, such as diminished screen visibility \cite{xiong2021augmented, erickson2022review} and social discomfort \cite{yuan2025head, goodman2019social}. Furthermore, users' experience and usage patterns may change as they gradually adapt to assistive technologies. These changes are crucial for the effective design and adoption of assistive technologies \cite{zolyomi2017technology}, yet cannot be captured in short-term, well-controlled lab studies. We fill this gap by tracking and analyzing PLV's use of an AR navigation application in uncontrolled outdoor scenarios through a diary study, deriving insights for more practical AI-powered AR systems for low vision.

\subsection{Field Study of Assistive Technologies for People with Disabilities}
\label{subsec:field_study_rw}
Many assistive technologies (ATs) for people with disabilities are evaluated in short-term lab studies, focusing on performance in predetermined tasks. 
However, such studies are limited in ecological validity \cite{bellucci2018research}, i.e., the extent to which findings generalize to real-world use. Lab studies that take place in ideal settings cannot reflect the dynamic and unpredictable nature of real-life environments \cite{montesano2010towards, bellucci2018research, hofstede2025field, federici2016providing}, where issues such as technical challenges (e.g., poor internet or GPS connectivity) \cite{pal2017agency, zhao2018face}, social acceptability \cite{singh2025social}, and users' changing needs and priorities \cite{phillips1993predictors, zolyomi2017technology} arise. 
Participants also tend to behave differently in artificial settings than in everyday life \cite{bellucci2018research}.
These limitations undermine the ecological validity of such AT research and can lead to abandonment of ATs in real-world use \cite{phillips1993predictors, petrie2018assistive}, highlighting the need for longitudinal field studies to understand real-world user experiences and challenges \cite{hofstede2025field, bellucci2018research}.

To better understand the real-world use of ATs, prior work has conducted field evaluations for people with different disabilities \cite{konig2022user, montesano2010towards, zhao2018face, xu2024imageexplorer, gonzalez2024investigating}. For example, K{\"o}nig et al. \cite{konig2022user} conducted a twelve-week home study on a tablet-based device assisting people with dementia to remember daily information and contact caregivers, and found that users' attitudes toward technology affected the device's usefulness.
For BLV people, Zhao et al. \cite{zhao2018face} deployed and evaluated a face recognition application through a seven-day diary study, revealing the real-world challenges faced by blind users when using this app, including difficulties aiming the camera and lack of knowledge about photo quality or phone status.
Gonzalez et al. \cite{gonzalez2024investigating} investigated the real-world deployment of an AI-powered scene description application. Through a two-week diary study, they characterized BLV users' usage scenarios and experiences with AI-generated scene description in daily life.
Xu et al. \cite{xu2024imageexplorer} explored how BLV users leveraged text-based and touch-based alt-text exploration in real-world usage by deploying a mobile application providing multi-layered image information via the two modalities. From real-world usage data across 12 months from 371 BLV users, they identified primary image categories users preferred with either modality, their usage patterns in touch-based exploration, and factors influencing user retention.

While these field evaluations have provided valuable insights into the real-world use of AT for the BLV community, they mainly focused on blind users and non-visual feedback. Despite the promise of visual augmentations for PLV, no research has investigated their ecological validity regarding PLV's real-world experiences, preferences, and interaction patterns outside of controlled lab environments. We address this gap by deploying a mobile AR system with low vision participants in real-world outdoor settings through a seven-day diary study. This study allowed us to understand the real-world use of visual augmentations and identify the unique environmental and social challenges \textit{in situ}.



\section{\system{}}
\label{sec:system}

To thoroughly understand PLV's real-world experiences and preferences with AR augmentations, we developed \system{}, a deployable mobile AR application that recognizes and enhances important outdoor objects in real time. The system integrates six augmentation designs inspired by prior work \cite{zhao2016cuesee, zhao2015foresee, lee2024cookar, fox2023using, maman2025enhancing, gopalakrishnan2025comparison}. By rendering object augmentations directly onto the phone's live camera feed on the screen, \system{} enables users to hold the phone to scan their surroundings and see an augmented world based on their object and augmentation selection.

To better understand what objects low vision users want to augment and how they want to augment different objects across scenarios, \system{} supports three levels of customization. First, \system{} recognizes a wide range of important outdoor objects, and users can flexibly select the objects they want to enhance. Second, \system{} provides two augmentation groups so that objects assigned to different groups can be augmented differently, allowing users to tailor augmentations to different object types and needs. Finally, \system{} provides six augmentations that enable users to customize their visual experience by selecting, combining, and adjusting augmentation parameters (e.g., color, contour thickness). We detail \system{}'s design and implementation below.


\subsection{Key Outdoor Objects to Augment}
\label{subsec:objects}

To enable PLV to select and augment objects relevant to their outdoor navigation, \system{} needs to recognize a wide range of such objects. Based on prior work on PLV's challenges during outdoor navigation \cite{riazi2016outdoor, cloutier2022topical, zeng2015survey, gamage2023blind, islam2024identifying},
we identified 21 object categories across four main challenges: (1) \textit{staying on walkable paths}, with objects indicating such paths and their boundaries: \textit{curb, sidewalk} \cite{islam2024identifying, gamage2023blind, zeng2015survey, riazi2016outdoor}; (2) \textit{avoiding obstacles}, with common obstacles that PLV may run into or trip over: \textit{rail track, fence, bench, sewer drain, fire hydrant, junction box, pole, traffic cone, trash can} \cite{islam2024identifying, gamage2023blind, riazi2016outdoor, cloutier2022topical}; (3) \textit{crossing streets}, with objects indicating where and when to cross: \textit{crosswalk, pedestrian light, vehicle traffic light, traffic sign} \cite{zeng2015survey, islam2024identifying, gamage2023blind, riazi2016outdoor, cloutier2022topical}; and (4) \textit{avoiding collisions with moving objects}, including \textit{pedestrian, bicycle, bus, car, motorcycle, truck} \cite{islam2024identifying, gamage2023blind, riazi2016outdoor}. Based on these categories, \system{} recognizes key objects selected by users in real time and renders their preferred visual augmentations. 

\subsection{Visual Augmentations}
\label{subsec:augmentation_design}
Inspired by the AR augmentation designs for low vision in prior work \cite{zhao2016cuesee, zhao2015foresee, lee2024cookar, fox2023using, maman2025enhancing, gopalakrishnan2025comparison}, we designed six visual augmentations,
including four \textbf{foreground effects} that directly augment specific objects: \textit{Contour Enhancement}, \textit{Solid Overlay}, \textit{Flashing}, and \textit{Brightness Adjustment}; and two \textbf{background effects} that reduce distractions: \textit{Background Darkening} and \textit{Color Removal}.

\begin{figure*}[htbp]
    \centering
    \includegraphics[width=\textwidth]{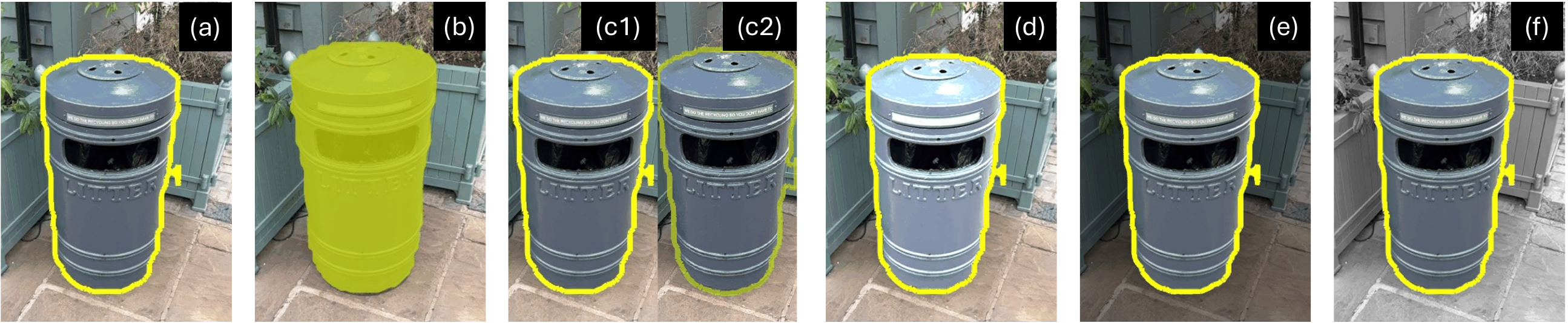}
    
  \caption{(a) \textit{Contour Enhancement}: a colored outline around the object's boundary.
  (b) \textit{Solid Overlay}: a semi-transparent colored overlay on the object region.
  (c1--c2) \textit{Flashing}: alternates a \textit{Contour Enhancement} or \textit{Solid Overlay} augmentation between 0\% and 100\% transparency every second.
  (d) \textit{Brightness Adjustment}: increases the object region's brightness to increase its contrast against the background.
  (e) \textit{Background Darkening}: dims the background to increase object contrast.
  (f) \textit{Color Removal}: renders the background in grayscale while keeping augmented objects in their original color.
    }
    \Description{
This figure shows the six augmentation designs of NavSight.
(a) The Contour Enhancement design. It augments object boundaries by rendering a colored outline around the object. 
(b) The Solid Overlay design. It renders a semi-transparent colored overlay on the object region. 
(c1-c2) The Flashing design. It generates a flashing effect to the Contour Enhancement or Solid Overlay augmentations by alternating the augmentation transparency between 0\% and 100\% every second.
(d) The Brightness Adjustment design. It increases the brightness of the object region to increase its contrast against the background.
(e) The Background Darkening design. It dims the background of the augmented objects to increase object contrast against the background.
(f) The Color Removal design. It renders the background in grayscale while keeping augmented objects in their original color.
    }
  \label{fig:design_list}
\end{figure*}

\textit{\textbf{Contour Enhancement}}. This design augments object boundaries by rendering a colored outline around the object, which increases the object visibility without obstructing it \cite{zhao2016cuesee, lee2024cookar, fox2023using, maman2025enhancing} (Figure~\ref{fig:design_list}a). Users can customize the contour color (yellow, green, red, blue, cyan, or white) and contour thickness (from 1\,px to 15\,px with 1\,px increments).

\textit{\textbf{Solid Overlay}}. This design renders a semi-transparent colored overlay on the object region \cite{lee2024cookar} (Figure~\ref{fig:design_list}b). Although it blocks some object details, this augmentation can be more visible for people with severe low vision. Users can customize the overlay color and transparency (from 0\% to 100\% with 1\% increments).  

\textit{\textbf{Flashing}}. This design applies a flashing effect to a \textit{Contour Enhancement} or \textit{Solid Overlay} augmentation by alternating its transparency between 0\% and 100\% every second (Figure~\ref{fig:design_list}c1--c2). This flashing effect can effectively attract low vision users' attention, especially for those with severe low vision \cite{zhao2016cuesee}.

\textit{\textbf{Brightness Adjustment}}. This design increases the brightness of the object region to increase its contrast with the background \cite{zhao2016cuesee, zhao2015foresee} (Figure~\ref{fig:design_list}d). The increment is adjustable between 0\% and 100\% with 1\% increments. 

\textit{\textbf{Background Darkening}}. This design dims the background to increase object contrast \cite{zhao2016cuesee, gopalakrishnan2025comparison} (Figure~\ref{fig:design_list}e). The dimming intensity is adjustable between 0\% and 100\% with 1\% increments.

\textit{\textbf{Color Removal}}. This augmentation renders the background in grayscale while keeping augmented objects in their original color to increase object visibility against the background \cite{zhao2016cuesee, gopalakrishnan2025comparison} (Figure~\ref{fig:design_list}f). This effect can be especially useful for low vision users with good color vision.

\subsection{Interaction Design}
\label{subsec:interaction_design}
As different object categories play distinct roles during outdoor navigation (e.g., walkable paths vs. collision hazards) \cite{islam2024identifying}, \system{} provides two \textit{Augmentation Groups} (\textit{Augmentation Group I} and \textit{Augmentation Group II}) to allow different augmentation selections. Users assign objects to each group and configure augmentation effects for each group. The group-based customization allows dedicated augmentations for certain objects while avoiding the high workload caused by per-object customization. Besides foreground augmentations for the two object groups, users can also select background effects (\textit{Background Darkening} and \textit{Color Removal}) to reduce distraction.

\begin{figure*}[htbp]
    \centering
    \includegraphics[width=\textwidth]{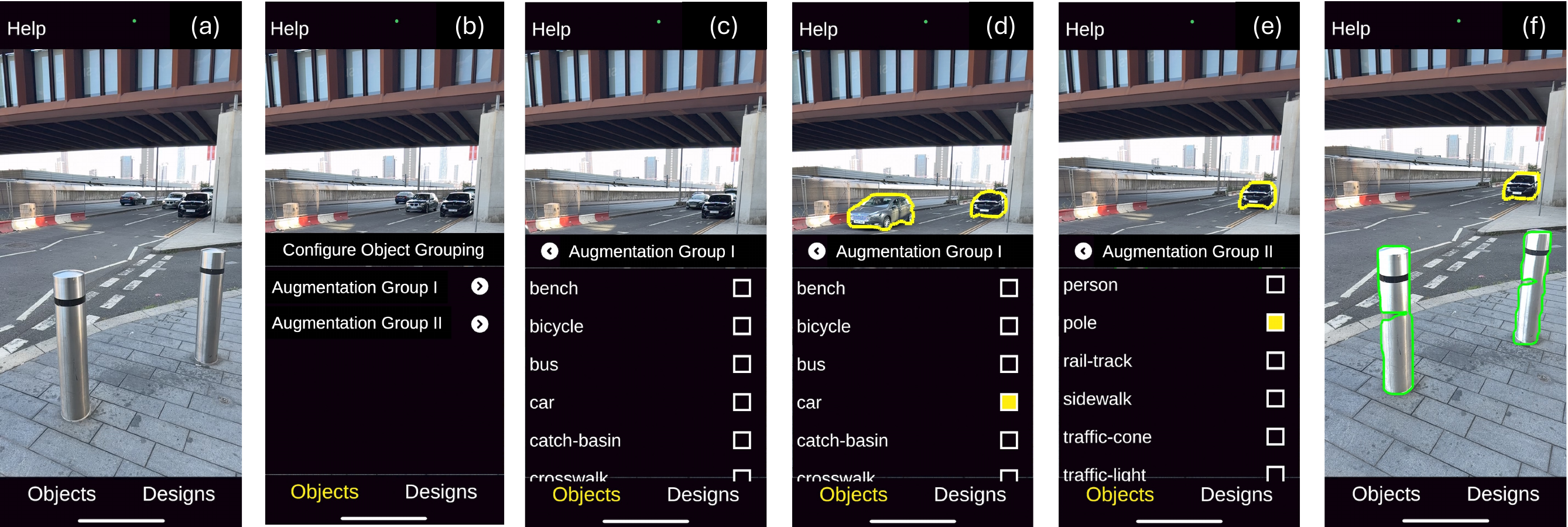}
    
  \caption{
  Interaction flow for assigning objects to augmentation groups.
  (a) Default view before configuration, with two poles and several cars in the scene, none augmented.  
(b) The ``Objects'' panel. Users can select object categories for each augmentation group by clicking the corresponding entry.
(c) Sub-panel for \textit{Augmentation Group I} listing object categories.  
(d) Assigning ``car'' to \textit{Augmentation Group I}. Cars are augmented with a yellow contour.  
(e) Assigning ``pole'' to \textit{Augmentation Group II}.
(f) Final view with cars in yellow contours (\textit{Augmentation Group I}) and poles in green contours (\textit{Augmentation Group II}).
}

\Description{
Six screenshots of NavSight showing how a user assigns object categories to augmentation groups. All six show the same street scene: two metal poles on a paved sidewalk in the foreground, a road with several parked cars in the middle distance, and an overpass above.
In (a), no objects are augmented, and a bottom bar shows two tabs, “Objects” and “Designs”.
In (b), the “Objects” tab is selected and a panel titled “Configure Object Grouping” covers the lower half of the screen, with one entry for each of the two augmentation groups.
In (c), the sub-panel for Augmentation Group I lists object categories in alphabetical order, including bench, bicycle, bus, car, crosswalk, and sewer drain, each with an unchecked box.
In (d), the box next to car is checked and the two cars in the scene are outlined in yellow.
In (e), the sub-panel for Augmentation Group II lists person, pole, rail track, sidewalk, traffic cone, and traffic light, with the box next to pole checked.
In (f), the panel is closed and both selections are visible in the scene, with a car outlined in yellow and the two poles outlined in green.
  }
  \label{fig:objects}
\end{figure*}

\begin{figure*}[htbp]
    \centering
    \includegraphics[width=\textwidth]{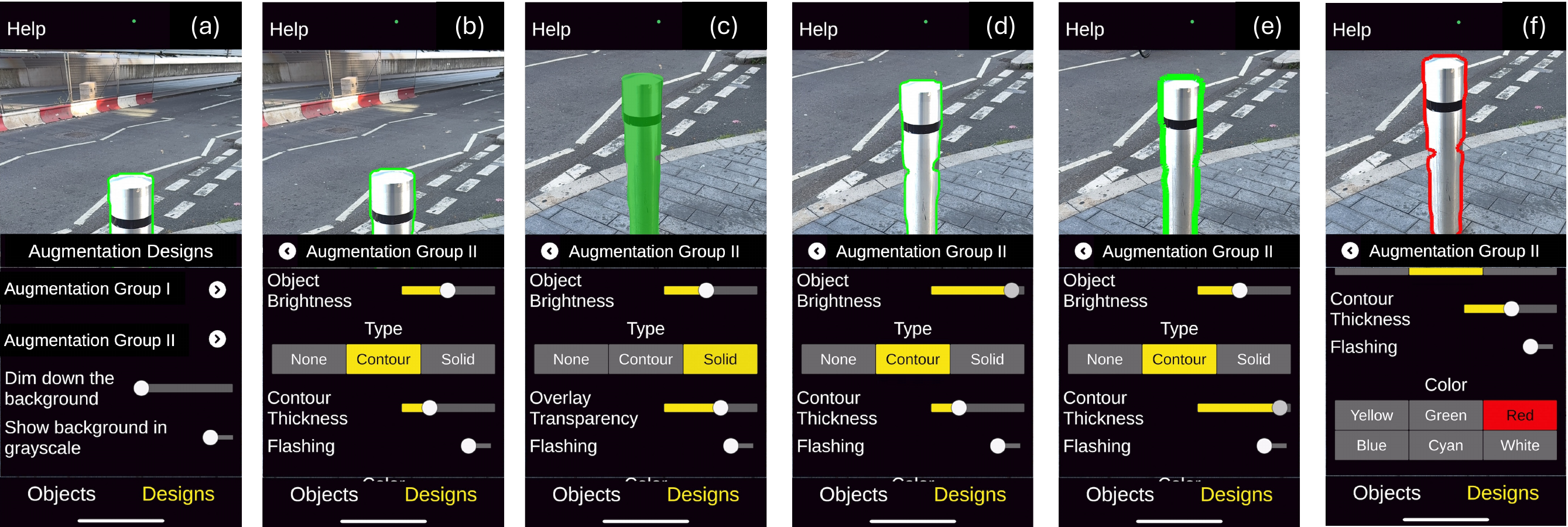}
    
  \caption{
  Interaction flow for adjusting augmentation designs.  
(a) The ``Designs'' panel with entries for the two augmentation groups and controls for \textit{Background Darkening} and \textit{Color Removal}.  
(b) Sub-panel for \textit{Augmentation Group II} with \textit{Contour Enhancement} selected.
(c) Switching to \textit{Solid Overlay}.  
(d) Applying \textit{Brightness Adjustment} to increase object brightness.  
(e) Customizing contour thickness.  
(f) Selecting the augmentation color from six options (yellow, green, red, blue, cyan, white). Red is shown as the example.
}
\Description{
Six screenshots of NavSight showing how a user adjusts the augmentation design for an object group. All six show the same scene, a metal pole on a paved sidewalk beside a road with white markings, with the “Designs” tab selected at the bottom.
In (a), a panel titled “Augmentation Designs” lists the two augmentation groups and two background controls, a slider to dim the background and a toggle to show the background in grayscale.
In (b), the sub-panel for Augmentation Group II shows an object brightness slider, a type selector set to Contour, a contour thickness slider, and a flashing toggle. The pole is outlined in a thin green contour.
In (c), the type selector is set to Solid and the contour thickness slider is replaced by an overlay transparency slider. The pole is filled with a semi-transparent green overlay.
In (d), the type is set back to Contour and the object brightness slider is moved to the right. The pole appears brighter within its green contour.
In (e), the contour thickness slider is moved to the right and the pole's green contour is noticeably thicker.
In (f), a color selector shows six options in two rows, yellow, green, and red on the first and blue, cyan, and white on the second, with red selected. The pole's contour is now red.
  }
  \label{fig:designs}
\end{figure*}

\system{}'s interface has two collapsible panels, ``Objects'' and ``Designs.'' In the ``Objects'' panel, users can assign objects to each augmentation group (Figure~\ref{fig:objects}); in the ``Designs'' panel, users can select and customize each group's foreground effects and apply the two background effects (Figure~\ref{fig:designs}). Users can also combine multiple foreground effects for each group, such as both \textit{Contour Enhancement} and \textit{Brightness Adjustment} (Figure~\ref{fig:designs}d).

\subsection{System Implementation}
\label{subsec:implementation}
We describe \system{}'s implementation details, including object recognition and prototype implementation, with technical evaluation on recognition models and system performance and overhead.


\paragraph{\textbf{Object Recognition: Model Fine-tuning and Evaluation}}
\label{par:object_recognition}
To fine-tune a model to recognize important outdoor objects (Section~\ref{subsec:objects}), we refined the Mapillary Vistas dataset \cite{neuhold2017mapillary}, a dataset of outdoor environments captured in diverse lighting and weather conditions, and split its 20,000 images 80/10/10 into training, validation, and test sets. 

We fine-tuned the YOLO11 model \cite{yolo11_ultralytics}, specifically its \textit{YOLO11l-seg} variant, on the refined dataset, resulting in the \textit{YOLO11l-seg-outdoor} model (150 epochs, batch size 64).
We evaluated \textit{YOLO11l-seg-outdoor} against the baseline \textit{YOLO11l-seg} model (pre-trained on MS-COCO \cite{lin2015mscoco} without fine-tuning) on the test set using three standard instance-segmentation metrics \cite{padilla2021comparative}: mAP50 and mAP75 (mean average precision at intersection-over-union (IoU) thresholds of 50\% and 75\%) and mAP (mean average precision averaged over IoU thresholds from 50\% to 95\% in 5\% increments). The fine-tuned model outperformed the baseline model on all three metrics (Table~\ref{tab:recognition_accuracy}).

We further evaluated the model's per-class recognition accuracy (Appendix~\ref{app:per-class}, Table~\ref{tab:per-class-accuracy}). The model was more accurate on objects with well-defined boundaries such as cars (AP=0.519) and pedestrians (AP=0.399), and less accurate on walkable paths such as sidewalk (AP=0.221), curb (AP=0.297), and crosswalk (AP=0.167), whose boundaries are less well-defined and whose appearance varies.
Further computing the false negative rate (FNR, the proportion of unaugmented important objects) and the false discovery rate (FDR, the proportion of augmentations that did not match an important object of the augmented category) \cite{padilla2021comparative}, we found both error types were more common for surfaces with ambiguous boundaries, such as crosswalks (FNR=0.791, FDR=0.284), and for objects that vary widely in appearance, such as benches (FNR=0.681, FDR=0.464) and sewer drains (FNR=0.697, FDR=0.400). In contrast, cars and pedestrians were both rarely missed and rarely augmented by mistake (car FNR=0.170, FDR=0.163; pedestrian FNR=0.251, FDR=0.243).
These per-class results are consistent with participants' reports that recognition of walkable paths was affected by real-world conditions (Section~\ref{subsubsec:env_factor_acc}) while recognition of objects like vehicles and pedestrians remained consistent. We discuss the implications of these results in Section~\ref{subsec:ai_in_the_wild}.

\begin{table}[t]
\centering

\caption{Recognition accuracy of the fine-tuned and baseline models. The fine-tuned model outperformed the baseline \textit{YOLO11l-seg} model on all three metrics in recognizing important outdoor objects for PLV.}
\label{tab:recognition_accuracy}
\begin{tabular}{p{3.5cm}rrr}
\toprule
  \textbf{Model} & \textbf{mAP50} & \textbf{mAP75} & \textbf{mAP} \\
\midrule
YOLO11l-seg (baseline) & 0.256 & 0.129 & 0.136 \\
YOLO11l-seg-outdoor & \textbf{0.588} & \textbf{0.305} & \textbf{0.318} \\
\bottomrule
\end{tabular}

\Description{This table compares the recognition accuracy between the baseline YOLO11l-seg model, pre-trained on MS-COCO without fine-tuning, and the fine-tuned model. This table reports three metrics: mAP50 and mAP75 (mean average precision at intersection-over-union (IoU) thresholds of 50\% and 75\%) and mAP (averaged over IoU thresholds from 50\% to 95\% in 5\% steps). The fine-tuned model scores higher than the baseline on all three metrics.}

\end{table}

\paragraph{\textbf{Prototype Implementation}}
\label{par:prototype_implementation}
We implemented \system{} on iOS because iOS devices are most accessible and most widely used by BLV people \cite{griffin2017survey}.
We used Unity 2022.3.20f1 for video capture, interface, and augmentation rendering, and Apple's Core ML \cite{apple_coreml} for on-device object inference. From the video stream, \system{} captures the video frame, detects objects with our fine-tuned model, and renders the selected augmentations in Unity. 

We added a screenshot feature to enable participants in the diary study (Section~\ref{sec:diary_study}) to report usage scenarios conveniently. Participants can double-tap the screen to capture the current camera frame with its augmentations. To mitigate privacy concerns, \system{} plays a shutter sound on capture to make the capture noticeable to bystanders \cite{denning2014situ}. 

\paragraph{\textbf{System Performance \& Overhead}}
\label{par:system_overhead}
We evaluated \system{}'s on-device latency and power consumption on an iPhone 13 by running \system{} continuously for 30 minutes. Each frame passes through three main processing stages: pre-processing (down-sampling the camera frame for inference, 3.3 ms), model inference (39 ms), and augmentation rendering (2.8 ms). Because inference runs on a background thread and rendering on the main thread, each frame incurs a 25 ms delay in passing the recognition result from one thread to the other. The resulting end-to-end latency is 70 ms per frame. Since \system{} processes frames in parallel, it reaches a throughput of around 25.6 FPS.

Following prior mobile AR work \cite{zhao2023multi}, we measured battery drain as the percentage of total capacity consumed over the 30-minute period. The battery drained by 20\%.
This power consumption came mainly from \system{}'s object recognition and augmentation: a camera-only baseline that rendered the live camera feed with no object recognition or augmentation drained around 5\% over the same period.

\section{Diary Study}
\label{sec:diary_study}
To understand the real-world experiences and usage patterns of \system{}, we conducted a seven-day diary study with 12 low vision participants. We seek to answer: (1) In what real-world scenarios do PLV use \system{}, and what are their usage patterns? (2) What strategies do PLV use to select and group objects and adjust augmentations across scenarios? (3) How do PLV make sense of and respond to \system{}'s recognition errors? and (4) What social and real-world usability challenges do PLV face when using \system{} in public?

\subsection{Participants}
We recruited 12 participants (three male, nine female) with ages ranging from 25 to 88 ($M=59.08$, $SD=19.55$) from local low-vision clinics and communities. Eligible participants had to:  
(1) be over 18 years old, (2) have low vision, 
and (3) own an iPhone (Section~\ref{par:prototype_implementation}).
Participants covered a broad range of visual conditions, including central vision loss (P12), peripheral vision loss (P2--P4, P8--P11), and severe low visual acuity (P2). Five participants (P2, P4, P9, P10, P12) were legally blind (i.e., visual acuity no better than 20/100 after best correction, or field of view narrower than 20 degrees \cite{aoaLegalBlindness}). 
Table~\ref{tab:demographics_diary} lists participants' demographics (i.e., age, gender), visual conditions, iPhone models used in the study, and number of days they used \system{}. Participants received \$200 upon study completion. This study was approved by our university IRB.

\begin{table*}[htbp]
\scriptsize
\centering
\caption{Participants' demographic information, visual conditions, their iPhone models for the diary study, and number of days they used \system{}. All participants used \system{} on at least five days ($M=7.92$, $SD=2.27$).
}
\label{tab:demographics_diary}
\begin{tabular}{>{\centering\arraybackslash}p{0.2cm}>{\centering\arraybackslash}p{0.3cm}>{\centering\arraybackslash}p{2.8cm}>{\centering\arraybackslash}p{0.2cm}>{\centering\arraybackslash}p{2.3cm}>{\centering\arraybackslash}p{2.4cm}>{\centering\arraybackslash}p{2cm}>{\centering\arraybackslash}p{1.5cm}>
{\centering\arraybackslash}p{1cm}}
\toprule
  \textbf{ID} & \textbf{Age/} \newline \textbf{Gender} &  \textbf{Diagnosis} & \textbf{Legally Blind} &  \textbf{Visual Acuity} &  \textbf{Field of View} &  \textbf{Other Visual Difficulties} & \textbf{iPhone Model} & \textbf{Days Used} \\

\hline
\multirow{2}{*}{P1} & \multirow{2}{*}{71/F} & Macular degeneration, & \multirow{2}{*}{N} & L: 20/20 & L: Intact & \multirow{2}{*}{Sensitive to light} & \multirow{2}{*}{iPhone 16} & \multirow{2}{*}{7} \\
& & astigmatism, cataract & & R: No functional vision & R: No functional vision & & &\\

\hline
\multirow{2}{*}{P2} & \multirow{2}{*}{62/F} & Spinal meningitis & \multirow{2}{*}{Y} & L: 20/2200 & Cannot see lower & \multirow{2}{*}{Sensitive to light} & \multirow{2}{*}{iPhone 14 Plus} & \multirow{2}{*}{8}\\
& & (optic nerve damage) & & R: 20/400 & half of vision & & &\\

\hline
\multirow{2}{*}{P3} & \multirow{2}{*}{88/M} & Glaucoma; macular & \multirow{2}{*}{N} & L: 20/30 & Cannot see upper & \multirow{2}{*}{N/A} & \multirow{2}{*}{iPhone 13 Pro Max} & \multirow{2}{*}{5}\\
& & degeneration on right eye & & R: 20/100 & half of vision & & &\\

\hline
\multirow{2}{*}{P4} & \multirow{2}{*}{48/M} & \multirow{2}{*}{Retinitis pigmentosa} & \multirow{2}{*}{Y} & L: 20/200 & L: $<10\degree$ & \multirow{2}{*}{Sensitive to light} & \multirow{2}{*}{iPhone 14 Pro Max} & \multirow{2}{*}{6}\\
& &  & & R: 20/200 & R: $<10\degree$ & & & \\

\hline
\multirow{2}{*}{P5} & \multirow{2}{*}{25/F} & \multirow{2}{*}{Antimetropia} & \multirow{2}{*}{N} & L: 20/200 & \multirow{2}{*}{Full} & \multirow{2}{*}{N/A} & \multirow{2}{*}{iPhone 14 Pro} & \multirow{2}{*}{12}\\
& &  & & R: 20/70 & & & &\\

\hline
\multirow{2}{*}{P6} & \multirow{2}{*}{73/F} & \multirow{2}{*}{Geographic atrophy} & \multirow{2}{*}{N} & L: 20/120 & \multirow{2}{*}{Full} & \multirow{2}{*}{N/A} & \multirow{2}{*}{iPhone 15 Pro} & \multirow{2}{*}{6}\\
& &  & & R: 20/90 & & & & \\

\hline
\multirow{2}{*}{P7} & \multirow{2}{*}{71/F} & Macular degeneration and & \multirow{2}{*}{N} & L: 20/150 & \multirow{2}{*}{Full} & \multirow{2}{*}{N/A} & \multirow{2}{*}{iPhone 12 Pro Max} & \multirow{2}{*}{8}\\
& & side effects of chemotherapy & & R: 20/300 & & & & \\

\hline
\multirow{2}{*}{P8} & \multirow{2}{*}{31/F} & \multirow{2}{*}{Retinitis pigmentosa} & \multirow{2}{*}{N} & L: 20/25 & L: $30\degree$ & \multirow{2}{*}{Sensitive to light} & \multirow{2}{*}{iPhone 13} & \multirow{2}{*}{11}\\
& & & & R: 20/25 & R: $30\degree$ & & & \\

\hline
\multirow{2}{*}{P9} & \multirow{2}{*}{54/F} & \multirow{2}{*}{Diabetic retinopathy} & \multirow{2}{*}{Y} & L: No functional vision & L: No functional vision & \multirow{2}{*}{Sensitive to light} & \multirow{2}{*}{iPhone 11} & \multirow{2}{*}{11}\\
& & & & R: 20/100 & R: Peripheral vision loss & & &\\

\hline
\multirow{2}{*}{P10} & \multirow{2}{*}{69/F} & \multirow{2}{*}{Albinism} & \multirow{2}{*}{Y} & L: 20/300 & \multirow{2}{*}{Peripheral vision loss} & \multirow{2}{*}{Sensitive to light} & \multirow{2}{*}{iPhone 13 Pro} & \multirow{2}{*}{6}\\
& & & & R: 20/250 & & & &\\

\hline
\multirow{2}{*}{P11} & \multirow{2}{*}{77/F} & \multirow{2}{*}{Glaucoma} & \multirow{2}{*}{N} & L: 20/20 & \multirow{2}{*}{Peripheral vision loss} & \multirow{2}{*}{N/A} & \multirow{2}{*}{iPhone 14} & \multirow{2}{*}{8}\\
& & & & R: 20/100 & & & &\\

\hline
\multirow{2}{*}{P12} & \multirow{2}{*}{40/M} & \multirow{2}{*}{Central scotoma} & \multirow{2}{*}{Y} & L: 20/400 & \multirow{2}{*}{Central vision loss} & \multirow{2}{*}{N/A} & \multirow{2}{*}{iPhone 14} & \multirow{2}{*}{7}\\
& & & & R: 20/300 & & & &\\


\bottomrule
\end{tabular}

\Description{This table lists out participant demographics in the diary study, their visual conditions, iPhone models for the study, and number of days they used NavSight. All participants used NavSight on at least five days ($M=7.92$, $SD=2.27$).}

\end{table*}

\subsection{Study Procedure}
The study consisted of three phases: an initial introduction, a week-long diary study, and a final interview.

\subsubsection{Initial Introduction}
We started the study with an in-person interview to collect participants' demographic information and visual condition. Then, we helped participants install \system{} on their phone using TestFlight \cite{appleTestFlightApple} and demonstrated key features, including selecting objects to augment, assigning objects to augmentation groups, adjusting visual augmentations for each object group, and capturing screenshots. We then guided participants to a nearby sidewalk and asked them to practice all features in \system{} until they were fully familiar with it. We also provided written instructions and encouraged participants to contact the research team at any time for troubleshooting.

\subsubsection{Diary Study}
During the diary study, participants were instructed to use \system{} on at least four days, across at least four scenarios, for a total of at least two hours to ensure sufficient usage across diverse real-world scenarios. 
We also asked participants to capture at least four screenshots in at least two scenarios to record representative or important usage moments for the final interview and data analysis. 
Participants filled out a daily survey (Table~\ref{tab:diary_questions}) to reflect on their app usage experiences, including whether they used \system{} that day and in what scenarios, and their experience in terms of perceived helpfulness, safety, recognition accuracy, distraction, and comfort using \system{} in public. The survey also asked whether they changed object group assignments or augmentations and why. We sent out an email each day at 5:00 PM to remind participants to fill out the survey. If participants did not use \system{} for sufficient time or days within a week, we extended the study until they met the requirements.

Besides the self-reported survey data, we also collected privacy-preserving usage logs from \system{}, including session duration for each launch and participants' object and augmentation selections per use. All usage logs and screenshots were stored on participants' phones and synced to a cloud server only accessible to the research team. Participants were notified and explicitly consented that the usage logs and screenshots would be uploaded automatically. Before further analysis on the screenshots, we removed screenshots that contained sensitive content and blurred identifiable elements (e.g., faces) to protect privacy.

\subsubsection{Final Interview}
After participants met all usage requirements, we conducted a semi-structured exit interview to discuss their experience with \system{}. Before the interview, we reviewed participants' usage logs, daily survey responses, and screenshots to identify noteworthy patterns, such as changes in object selection, grouping, or augmentations; instances where they found \system{} unhelpful, distracting, inaccurate, or socially uncomfortable to use; and unexpected usage scenarios beyond outdoor navigation. 

During the interview, we used participants' screenshots and survey responses as probes and asked them to: (1) describe scenarios in which they chose to use \system{} and why; (2) explain their rationales for assigning objects to augmentation groups and selecting corresponding augmentations across scenarios; (3) reflect on their usage experience in terms of perceived helpfulness, safety, and comfort using \system{} in public, and scenarios in which they would be willing to use \system{}; and (4) discuss usability issues that affected their experience, such as form-factor preferences, the impact of recognition errors, and the strategies they used to cope with such errors.

\begin{table*}[htbp]
\small
\centering
\caption{Daily survey questions in the diary study.}
\label{tab:diary_questions}
\begin{tabular}{>{\centering\arraybackslash}p{0.5cm}>{\arraybackslash}p{14cm}}
\toprule

\textbf{Q1} & Did you use \system{} today? (If no, will jump to Q11) \\

\hline
\textbf{Q2} & In what scenarios did you use \system{} today? \\

\hline
\textbf{Q3(a)} & How helpful do you feel \system{} is today? (1=``Not helpful at all'', 5=``Extremely helpful'') \\

\hline
\textbf{Q3(b)} & Why did you give this rating? \\

\hline
\textbf{Q4(a)} & How safe do you feel navigating outdoors with \system{} today? (1=``Not safe at all'', 5=``Extremely safe'') \\

\hline
\textbf{Q4(b)} & Why did you give this rating? \\

\hline
\textbf{Q5(a)} & How accurate do you feel \system{} is in recognizing objects? (1=``Not accurate at all'', 5=``Extremely accurate'') \\

\hline
\textbf{Q5(b)} & Why did you give this rating? \\

\hline
\textbf{Q6(a)} & How distracting do you feel when using \system{} today? (1=``Not distracting at all'', 5=``Extremely distracting'') \\

\hline
\textbf{Q6(b)} & Why did you give this rating? \\

\hline
\textbf{Q7(a)} & How comfortable do you feel using \system{} in public settings? (1=``Not comfortable at all'', 5=``Extremely comfortable'') \\

\hline
\textbf{Q7(b)} & Why did you give this rating? \\

\hline
\textbf{Q8} & Did you change the augmentation group of objects today when using the app? Why? \\

\hline
\textbf{Q9} & Did you change the augmentation designs today when using \system{}? Why? \\

\hline
\textbf{Q10} & Do you have any comments, questions, or suggestions related to \system{}? \\

\hline
\textbf{Q11} & (Only when users did not use \system{} today) What is the reason you didn't use the app today? \\

\bottomrule
\end{tabular}

\Description{Daily survey questions in the diary study.}

\end{table*}

\subsection{Data Analysis}
\label{subsec:data_analysis}
All interviews were audio recorded and transcribed using an automatic transcription service. We analyzed the daily survey entries and transcripts using thematic analysis \cite{braun2006using, clarke2017thematic}. Two researchers independently open-coded three participants' data (25\% of the data) and discussed them to develop an initial codebook. They then independently coded the rest of the data, periodically discussing and adding new codes upon agreement. Finally, we developed themes and sub-themes by clustering relevant codes using axial coding and affinity diagrams \cite{terry2017thematic}. We also analyzed participants' subjective ratings in the daily surveys through descriptive statistical analysis (e.g., median, range). 

We further analyzed the application usage logs across participants. We computed the number of days each participant used \system{} and their usage duration. Because participants may switch between applications or use their phones intermittently during navigation, we grouped consecutive usage periods with short intervals into one usage session. Two usage periods needed to be at least five minutes apart to be counted as distinct sessions \cite{zhao2017effect}. We excluded sessions shorter than ten seconds. We also tracked the augmented objects, object grouping, and augmentation selections across days to analyze how their usage evolved.
\section{Findings}
\label{sec:findings}
Our study revealed how PLV used a mobile AR visual augmentation aid (i.e., \system{}) in real-world outdoor environments. We report their usage patterns and scenarios, perception and attention allocation changes through \system{}, object and augmentation selections, error perception and coping strategies, and usability challenges in real-world use.


\subsection{Usage Patterns and Scenarios}
\label{subsec:usage_pattern_scenarios}
We first report participants' usage patterns across the diary study, including frequency, duration, and common and novel scenarios. 
\subsubsection{Frequency and Duration of Use}
\label{subsubsec:frequency_duration}
\begin{figure*}[tbp]
    \centering
    \begin{subfigure}[b]{0.49\textwidth}
        \centering
        \includegraphics[width=\textwidth]{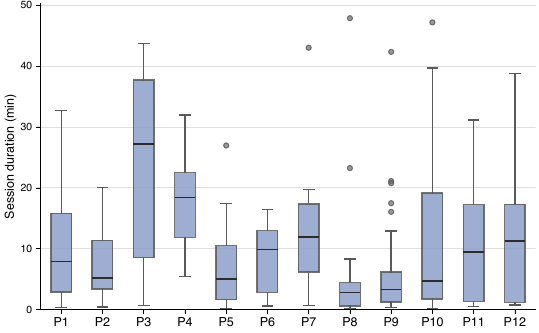}
        \caption{}
        \label{subfig:session_duration}
    \end{subfigure}
    \hfill
    \begin{subfigure}[b]{0.49\textwidth}
        \centering
        \includegraphics[width=\textwidth]{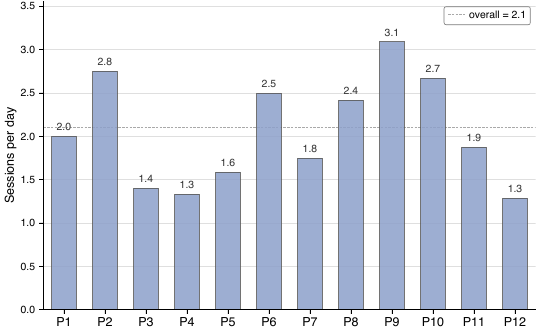}
        \caption{}
        \label{subfig:sessions_per_day}
    \end{subfigure}

  \caption{
  Participants' usage of \system{} across the study.
  (a) Distribution of session duration for each participant. Each box spans the 25th to 75th percentiles of that participant's session durations, with a bold line marking the median. Whiskers extend to the most extreme point within 1.5 times the interquartile range, and dots indicate sessions beyond the whiskers.
  (b) Mean number of sessions per day for each participant, with the dashed line marking the overall mean of 2.1 sessions per day.
  Together, the two panels show that the number of sessions per day was negatively correlated with session duration. P8 and P9 had more sessions per day than the overall mean and the shortest sessions, whereas P3 and P4 had fewer sessions per day than the overall mean and the longest sessions.
  }
  \Description{
Two charts describing how participants used NavSight. The first is a box plot of session duration in minutes for participants P1 through P12. Median session duration ranges from about 3 minutes for P8 and P9 to about 27 minutes for P3 and 18 minutes for P4, with most other participants falling between 5 and 12 minutes. The spread also differs: P8 and P9 have narrow boxes concentrated below 7 minutes, while P3, P10, P11, and P12 have wide boxes spanning roughly 1 to 38 minutes. Five participants have sessions beyond their whiskers, the longest being about 48 minutes for P8, 47 for P10, 43 for P7, and 42 for P9. The second is a bar chart of the mean number of sessions per day for the same participants, ranging from 1.3 for P4 and P12 to 3.1 for P9, with P2 at 2.8 and P10 at 2.7 also above the overall mean of 2.1. The two panels are inversely related: participants with more sessions per day tended to have shorter sessions.
  }
  \label{fig:usage_patterns}
\end{figure*}

All participants used \system{} regularly for five to 12 days ($M = 7.92$, $SD = 2.27$), with one to three sessions of usage per day and each session lasting for 9.3 minutes on average, indicating considerable amounts of daily usage (Figure~\ref{fig:usage_patterns}). P2 continued using it after the study ended. 
Usage patterns varied across participants, with mean session durations ranging from around five minutes (P8) to over 20 minutes (P3) (Figure~\ref{subfig:session_duration}). We observed a negative correlation between the number of sessions per day and mean session duration (Pearson $r=-0.69$). P8 and P9 used \system{} in frequent, short sessions, with 2--5 sessions per day
and 5--7 minutes per session on average. In contrast, P3 and P4 had fewer but longer sessions, with 1--2 sessions per day and 17--24 minutes per session on average.
P3 and P4 also adopted different usage patterns in long sessions. P4 usually had the app running continuously during each session (e.g., one continuous 23.5-minute session).
P3 instead never used \system{} while walking, considering it unsafe. He stopped at intersections, driveways, and obstacles, opened \system{} to scan and verify what he saw, then closed it and walked with his own vision, launching it 28 times within one 36-minute session.

Participants' usage also varied across scenarios. For example, P9 typically used \system{} in short sessions of around five minutes at parking lots, checking her surroundings for cars and sidewalks before walking into the building. However, when visiting the zoo, she used it for seven longer sessions averaging 11 minutes, including one continuous 42-minute session, to check the queue size (Section~\ref{subsubsec:reappropriation}).

\subsubsection{Common Usage Scenarios for Navigation}
\label{subsubsec:common_scenarios}
Each participant used \system{} in at least four different scenarios, such as sidewalks (12/12), driveways (P1, P7, P10), parking lots (7/12; e.g., P5, P8), street crossings (6/12; e.g., P4, P8), and local parks (P6, P9, P10, P11).
Across these scenarios, the most common goal was to \textit{locate and remain on walkable paths} (12/12; Figure~\ref{fig:use_cases}a). Participants relied on \system{} for this across different paths, including structured sidewalks (12/12) and less structured surfaces such as roads without sidewalks (P7) and park trails (P6, P9, P10, P11). P1 also used \system{} to find walkable areas indoors (e.g., aisles of a shopping mall).

Participants also used \system{} to \textit{spot and avoid obstacles} (Figure~\ref{fig:use_cases}b), including vehicles on roads and in driveways (9/12; e.g., P1, P5), incoming pedestrians and cyclists (6/12; e.g., P3, P8), changes in ground elevation (10/12; all except P1, P12) such as curbs and curb cuts, and static obstacles such as poles (8/12; e.g., P3, P9). Four participants (P2, P6, P7, P10) also used \system{} to notice construction detours ahead of time to adjust their route.

\system{} also helped participants \textit{assess vehicle traffic}. Seven participants (e.g., P1, P6) used \system{} in parking lots to detect backing vehicles. Six participants (e.g., P2, P8) used it at intersections before stepping into the street (Figure~\ref{fig:use_cases}c), such as checking right-turning vehicles (P5).
However, participants showed divergent opinions on using \system{} during street crossing. P1 and P8 preferred using it to stay within crosswalks by highlighting their boundary, while P2 relied on her own vision and decided against using \system{}: \textit{``I use the app to know when I can cross the street, and [when crossing] I felt the app was distracting because I was trying to pay attention to the app and not what I was doing.''}

Finally, participants used \system{} to \textit{perceive the scene and locate specific objects} (P2, P5, P8, P12). P2 and P8 augmented the bus to see it approaching so they could get ready without having to hurry (Figure~\ref{fig:use_cases}d). P12 similarly used \system{} to check whether his ride-share had arrived, and P5 augmented street signs to locate them in unfamiliar neighborhoods to know which streets she was on. 

\subsubsection{Usage Scenarios Beyond Navigation}
\label{subsubsec:reappropriation}
\begin{figure*}[tbp]
    \centering
    \includegraphics[width=\textwidth]{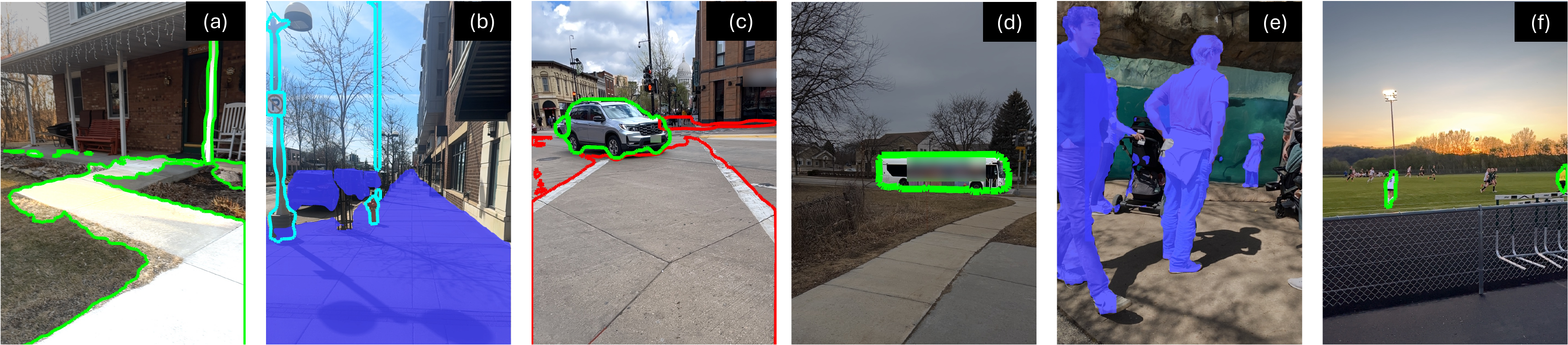}
    
  \caption{Example screenshots of participants' usage scenarios of \system{}.
  (a) Locate and remain on walkable paths, with the sidewalk and poles augmented with green contours and increased brightness;
  (b) Spot and avoid obstacles, with the sidewalk augmented with a blue solid overlay and obstacles (poles and a fire hydrant) with cyan contours;
  (c) Assess vehicle traffic, with cars augmented with green contours and crosswalks and sidewalks with red contours;
  (d) Observe the scene to locate a bus, with only the bus augmented with green contours;
  (e) Measure queue size, with people augmented with semi-transparent blue solid overlays;
  (f) Follow a soccer game, with players augmented with green contours.
  }
  \Description{This figure (with six subfigures a to f) shows example screenshots of participants' usage scenarios of \system{}.
  (a) Participants used it to locate and remain on walkable paths, with the sidewalk and poles augmented with green contours and increased brightness;
  (b) Participants used it to spot and avoid obstacles, with the sidewalk augmented with blue overlays and obstacles (poles and a fire hydrant) augmented with cyan contours;
  (c) Participants used it to assess traffic condition, with cars augmented with green contours and crosswalks and sidewalks with red contours;
  (d) Participants used it to watch for a bus, with only buses augmented with green contours and no other objects augmented;
  (e) Participants used it to measure queue size, with people augmented with semi-transparent blue solid overlays;
  (f) Participants used it to watch a soccer game, with players augmented with green contours.
  }
  \label{fig:use_cases}
\end{figure*}

Two participants (P8, P9) used \system{}'s augmentations to understand a scene for purposes beyond navigation.
P9 used it to estimate queue size (Figure~\ref{fig:use_cases}e). As she described using it in a zoo: ``\textit{It's easy to use [\system{}] to see how many people are [standing] against [an animal] [...] and [to see] when they leave, because the blue [augmentation on people] starts moving.}''
P8 checked the status of a soccer game (Figure~\ref{fig:use_cases}f) by augmenting and tracking the players. From the augmentations, she could quickly locate them to better understand the game. 
These cases suggest that \system{} can support broader \textit{in-situ} scene understanding beyond navigation.



\begin{figure*}[tbp]
    \centering
    \begin{subfigure}[b]{0.49\textwidth}
        \centering
        \includegraphics[width=\textwidth]{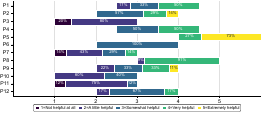}
        \caption{}
        \label{subfig:helpful}
    \end{subfigure}
    \hfill
    \begin{subfigure}[b]{0.49\textwidth}
        \centering
        \includegraphics[width=\textwidth]{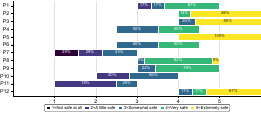}
        \caption{}
        \label{subfig:safe}
    \end{subfigure}
    
  \caption{
  Distribution of each participant's daily ratings of (a) perceived helpfulness and (b) perceived safety. Each bar represents one participant and is centered on that participant's median rating, with colored segments showing the proportion of that participant's ratings at each level, from 1 (not helpful at all or not safe at all) to 5 (extremely helpful or extremely safe). Higher ratings indicate that participants found \system{} more helpful and felt safer using it.
  }
  \Description{
  Two horizontal diverging bar charts, one for perceived helpfulness and one for perceived safety, each with one bar per participant from P1 to P12. Each bar is centered on that participant's median rating, and colored segments show the percentage of that participant's ratings at each of the five levels, from 1 (not helpful at all or not safe at all) to 5 (extremely helpful or extremely safe).
  In the helpfulness chart, P5 is the most positive, with 73 percent of her ratings at 5 and 27 percent at 4. P8 gave 4 on 91 percent of her days. P3 is the most negative, with 80 percent at 2 and 20 percent at 1, followed by P11 with 75 percent at 2, P10 with 60 percent at 2 and 40 percent at 3, and P7 with 43 percent at 2 and 14 percent at 1. P6 gave a rating of 3 on all of her days. P7 and P9 each used four of the five rating levels, more than any other participant.
  In the safety chart, P5 gave 5 on all of her days, and P2, P3, and P12 gave 5 on most days. P7 is the most negative, with 29 percent at 1, 29 percent at 2, and 43 percent at 3, followed by P11 with 75 percent at 2 and 25 percent at 3, and P10 with 40 percent at 2 and 60 percent at 3. All other participants have a median of 3 or above.
  }
  \label{fig:helpful_safe_bars}
\end{figure*}

    

\subsection{Enhanced Environment Perception Shaped by \system{}} 
\label{subsec:perception}
Most participants found \system{} helpful for navigation and felt safe using it, with eight rating it at least somewhat helpful and at least somewhat safe (median $\geq 3$ on both), because it improved their perception of the outdoor environment. Ratings differed across participants (Figure~\ref{subfig:helpful}, Figure~\ref{subfig:safe}), with median helpfulness ranging from 2.00 (P3, P7, P10, P11) to 5.00 (P5) ($SD = 0.93$), and median safety from 2.00 (P7, P11) to 5.00 (P2, P3, P5, P12) ($SD=1.09$). Although participants' perceived safety varied, no participant reported a fall or injury while using \system{}. Participants also raised concerns about \system{}, which we report in later sections.

Across participants, we identified four ways \system{} reshaped how they perceived the outdoor environment: it simplified the scene into walkable and non-walkable regions, made tripping hazards easier to notice, improved perception of moving objects, and extended visual reach. Together, these opened up new exploration opportunities, making some participants more willing to walk in unfamiliar places. 
We elaborate on the perception improvements below.

\subsubsection{Simplifying the Scene into Walkable and Non-walkable Regions}
\label{subsubsec:simplify_walk}
Navigating outdoors required participants to track many things at once, which they found distracting. As P2 explained: ``\textit{I am more distracted walking without the app because I have to be looking for everything: cars, people, obstacles, etc.}''
All participants reported that \system{} helped them interpret the \textit{walkability} of the environment (i.e., whether a region is safe to walk or not) by making walkable regions and their boundaries visually salient. Rather than searching for surface edges in the real world, participants used the augmentations to stay within walkable regions, including sidewalks (9/12; e.g., P1, P8) and crosswalks (P1, P8, P12). 
As P1 explained: ``\textit{[The sidewalk augmentation] tells me where I'm going. I know that there are two green lines along the side, so that's where I'm supposed to stay ... I knew [if] I just stayed in the middle of the green lines, I didn't fall over.}'' She rated \system{} very helpful (rating of 4) and very safe (4) when it recognized sidewalks, curbs, and crosswalks. P8 also found the augmentation of the walkable path reduced her cognitive load
: ``\textit{[NavSight] took away the extra thought process of: am I on the sidewalk, am I in between the sidewalk, am I out of people's way?}''
P8 also emphasized that in rainy weather, the curb became visually similar to the sidewalk and she relied on \system{} to distinguish between the two surfaces to avoid falling.

\subsubsection{Perceiving Tripping Hazards through Segmentation Cues}
\label{subsubsec:seg_cues}
Ten participants (e.g., P5, P11) reported that \system{} helped them notice tripping hazards along their path, 
like curbs, sewer drains, and rail tracks.
Interestingly, \system{} also enabled participants to detect hazards that it could not recognize. Because the sidewalk augmentation covered only the sidewalk surface, obstacles on it were excluded and appeared as holes within the augmentation (e.g., cracks or metal cords). Three participants (P2, P3, P8) used these ``holes'' as cues to notice and avoid the hazards. This suggests that segmentation-based augmentations can support hazard detection---by revealing the hazard as a gap in the augmentation---even when the hazard is not explicitly recognized.

\subsubsection{Improving Perception of Moving Objects}
\label{subsubsec:moving_objects}
Seven participants (e.g., P5, P12) reported that \system{} helped them notice approaching vehicles, pedestrians, and cyclists
, so they could respond in time.  
For example, six participants (e.g., P1, P4) noticed pedestrians earlier so that they could move to the side (P2), and P5 and P8 spotted incoming cyclists to avoid collision. In parking lots, P8 rated \system{} very helpful (4): ``\textit{[NavSight] made sure that I could highlight the cars so I could stop and let them go through.}'' 

In addition to holding \system{} in front of them, five participants (e.g., P2, P6) also aimed it toward directions from which traffic could come to scan for vehicles. P2 described scanning when crossing driveways: ``\textit{There are three driveways that I have to cross ... I would walk down and the driveways were on my right, so I'd pan the app to the right to see if there were any cars coming out, and then I would pan it to the left to see if there were any cars turning in off of the road.}'' In one entry, she reported catching a car turning into a driveway right in front of her that she would not have seen otherwise, and rated \system{} very helpful (4) and extremely safe (5).

\subsubsection{Extending Visual Reach}
\label{subsubsec:visual_reach}
Nine participants (e.g., P2, P9) described how \system{} helped them plan ahead during outdoor navigation by extending their visual reach and allowing them to notice relevant objects earlier.
For example, P7 crossed the road to avoid a construction barrier augmented with a contour, since she saw it on the app earlier than she could have seen it without the app. Although she rated \system{}'s helpfulness low overall (median=2) due to \system{} not recognizing uneven terrain on rural trails where she often walked, she rated it very helpful (4) on that day. 

However, the perceived helpfulness depended on whether the reach of \system{} could go beyond participants' own vision. Five participants (e.g., P4, P12) reported that \system{} did not recognize objects far enough to be useful. 
For P3 and P6, this range fell within their own viewing ability, making \system{} less useful. As a result, P6 rated it somewhat helpful (3) every day, and P3 rated it a little helpful (2) or below on every day, noting that obstacles ``were nicely outlined'' but only after he had already seen them.

\subsubsection{Opening Up Exploration Opportunities}
Building on \system{}'s improvements in how they perceived the environment, five participants (e.g., P2, P8) described feeling more confident and independent when navigating outdoors alone. P8 reported being more willing to walk alone in places she would otherwise have avoided, opening up new opportunities for exploration. As she said: ``\textit{I liked using it [\system{}] in different locations because that opens up the accessibility and not feeling limited to only certain routes and places that I can walk, that I feel the most safe. It opens up like, oh, I can take a little different route to a different store if I need to.}'' 


\begin{figure*}[tbp]
    \centering
    \begin{subfigure}[b]{0.49\textwidth}
        \centering
        \includegraphics[width=\textwidth]{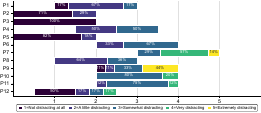}
        \caption{}
        \label{subfig:distraction}
    \end{subfigure}
    \hfill
    \begin{subfigure}[b]{0.49\textwidth}
        \centering
        \includegraphics[width=\textwidth]{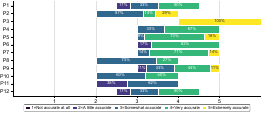}
        \caption{}
        \label{subfig:accuracy}
    \end{subfigure}
    
  \caption{
  Distribution of each participant's daily ratings of (a) perceived distraction and (b) perceived accuracy, each from 1 to 5. Each bar is centered on that participant's median rating. Higher accuracy ratings indicate that participants found \system{}'s recognition more accurate, whereas higher distraction ratings indicate that participants found \system{} more distracting.
  }
  \Description{
  Two horizontal diverging bar charts, one for perceived distraction and one for perceived accuracy, each with one bar per participant from P1 to P12. Each bar is centered on that participant's median rating, and colored segments show the percentage of that participant's ratings at each of the five levels.
  In the distraction chart, ratings run from 1 (not distracting at all) to 5 (extremely distracting), and lower ratings indicate less distraction. P3 gave a rating of 1 on all of his days, and P5 and P2 gave 1 on 82 and 71 percent of their days respectively. P7 is the most distracted, with 57 percent of her ratings at 4 and 14 percent at 5, and is the only participant with a median above 3. P9 also rated high, with 44 percent at 5, and used four of the five rating levels, more than any other participant. P10 and P11 gave 3 on 80 and 75 percent of their days.
  In the accuracy chart, ratings run from 1 (not accurate at all) to 5 (extremely accurate). P3 gave a rating of 5 on all of his days, the highest of any participant. P5, P6, and P7 rated 4 on most of their days, at 73, 83, and 71 percent respectively. P11 is the lowest, with 38 percent of her ratings at 2 and 62 percent at 3. P2, P8, P10, and P11 have medians of 3. No participant rated NavSight not accurate at all (1) on any day.
  }
  \label{fig:distraction_accuracy}
\end{figure*}

\subsection{Attention \& Effort Competition Caused by \system{}}
\label{subsec:attention_competition}
Participants' median perceived distraction ranged from 1.00 (P2, P3, P5) to 4.00 (P7) (Figure~\ref{subfig:distraction}), with nine participants occasionally reporting \system{} to be somewhat to very distracting and P7 and P9 sometimes finding it extremely distracting. 
Below, we report how \system{} competed for participants' visual attention and for the hands they needed while navigating, how participants adapted to the divided attention over time,
and their suggestions for wearable displays to address both competitions.


\subsubsection{Competing for Visual Attention}
\label{subsubsec:distraction}
Eight participants (e.g., P6, P11) reported that \system{} divided their attention between the phone screen and their surroundings. On her first day, P11 rated \system{} very distracting (4) and only a little safe (2), because locating the objects on the screen took her attention from where she was stepping.
As she explained: ``\textit{when I'm watching the phone, I'm not watching where my feet are going.}'' P12 similarly described focusing on the phone as ``compartmentalizing your world into a five- or six-inch screen.''

This divided attention lowered some participants' sense of safety, since watching the screen reduced their awareness of the hazards \system{} did not augment (P6, P10, P11). 
As P6 explained: ``\textit{even if hazards are identified, there are still other hazards that are missed by the app, making it uncomfortable to walk with my head down watching the screen.}'' P11 added that focusing on the phone was especially risky on bike paths, where she could not see approaching bicycles from the side. 

We also found that the perceived distraction depended on the activity. The more an activity required attention to the surroundings, such as crossing a street or walking with others, the more distracting and less safe it felt to navigate with \system{} (P4, P12). For example, P12 rated it not distracting at all (1) and extremely safe (5) when stationary, somewhat distracting (3) and very safe (4) while walking, and very distracting (4) and only somewhat safe (3) while crossing a street. As he explained, at a crossing he felt it was ``really distracting to try to look at your phone when also trying to be aware of cars around you.'' He therefore preferred to watch the traffic himself rather than use \system{}.

%
%


\subsubsection{Competing for the Hands}
\label{subsubsec:compete_hands}
Using \system{} also occupied a hand that participants needed while navigating (5/12), such as when keeping balance on uneven terrain (P4, P7), grabbing a handrail on stairs (P7), holding a cane (P11, P12), or carrying items (P1, P4). Managing the phone during these tasks added to participants' distraction and reduced their sense of safety. As P11 noted, with the cane in one hand and the phone in the other, ``\textit{if I fall, I don't have a free hand to stop me.}''
P4 rated \system{} somewhat distracting (3) and somewhat safe (3) when his hands were occupied, such as when holding a dog leash or groceries, while he rated it a little distracting (2) and very safe (4) when both hands were free.

%
%

\subsubsection{Managing the Divided Attention}
\label{subsubsec:manage_attention}
Interestingly, we found some participants adapted to this divided attention over the one-week usage, with five participants (e.g., P5, P11) demonstrating a decline in their rated distraction. P2's and P5's ratings declined within the first two days and stayed at not distracting at all (1) on all later days, with P5 reporting on her fourth day: ``\textit{I've gotten used to the app and do not find it distracting.}''
P11's ratings gradually declined, from very distracting (4) on her first day to a little distracting (2) on her last, reporting: ``\textit{I'm getting more comfortable walking with the phone in my field of vision.}''
P3 instead managed the attention competition by looking at \system{} only when standing still to check the scene and putting it away to walk, rating it not distracting at all (1) across all five usage days.

To diminish the attention division, six participants (e.g., P3, P8) suggested deploying \system{} on wearable displays (e.g., smart glasses). Three (P3, P11, P12) wanted the augmentations shown directly in their line of sight so they would not need to look away from their surroundings to a separate screen. As P3 explained: ``\textit{[wearable displays would provide] real-time feedback where I didn't have to take away my attention from the walking and look at an external device.}''
Five participants (e.g., P4, P8) also wanted a wearable display so they could have both hands free for navigating. 

\subsection{Interpretation of AI Errors}
\label{subsec:accuracy}
Participants rated \system{}'s accuracy positively overall, with the median rating of all twelve participants at least somewhat accurate (3) and no participant rating it not accurate at all (1) on any day (Figure~\ref{subfig:accuracy}). Their median ratings ranged from 3.00 (P2, P8, P10, P11) to 5.00 (P3). However, participants still occasionally encountered recognition errors, and five participants rated \system{} a little accurate (2) on at least one day.
Below, we report the types of errors participants encountered, the environmental factors affecting recognition, participants' attitudes toward different errors, and how they interpreted and responded to these errors.

    

\subsubsection{Recognition Error Types}
\label{subsubsec:misrecognition}
Participants described four common types of recognition errors: 
(1) \textbf{Segmentation Inaccuracy}: Seven participants (e.g., P8, P10) noted that the sidewalk augmentation sometimes appeared ``squiggly'' and changed shape over time due to inaccurate boundaries, which caused confusion. 
(2) \textbf{Temporal Instability}: Five participants (e.g., P4, P12) reported seeing flickering augmentations caused by inconsistent recognition results between consecutive frames (e.g., an object was recognized in one frame but not recognized in the next frame). They found the flickering augmentations distracting and visually irritating.
(3) \textbf{False Positives on Out-of-List Objects}: Five participants (e.g., P4, P11) observed \system{} augmenting objects outside the supported categories (Section~\ref{subsec:objects}) when they were visually similar to in-list objects (e.g., a car stopper in a parking lot being mistaken for a curb; a handrail being mistaken for a fence).
(4) \textbf{Missed Objects at Uncommon Orientations}:
Three participants (P2, P7, P12) reported that \system{} could not recognize objects at uncommon orientations (i.e., atypical poses and angles). P2 found that \system{} recognized traffic cones only when they were upright, and P12 similarly noted that \system{} recognized a bicycle standing upright but not once it had fallen over. P7 also noted that a bicycle approaching head-on was ``too thin'' to be recognized.

\subsubsection{Environmental Factors Affecting AI Accuracy}
\label{subsubsec:env_factor_acc}
\begin{figure*}[tbp]
    \centering
    \includegraphics[width=0.6\textwidth]{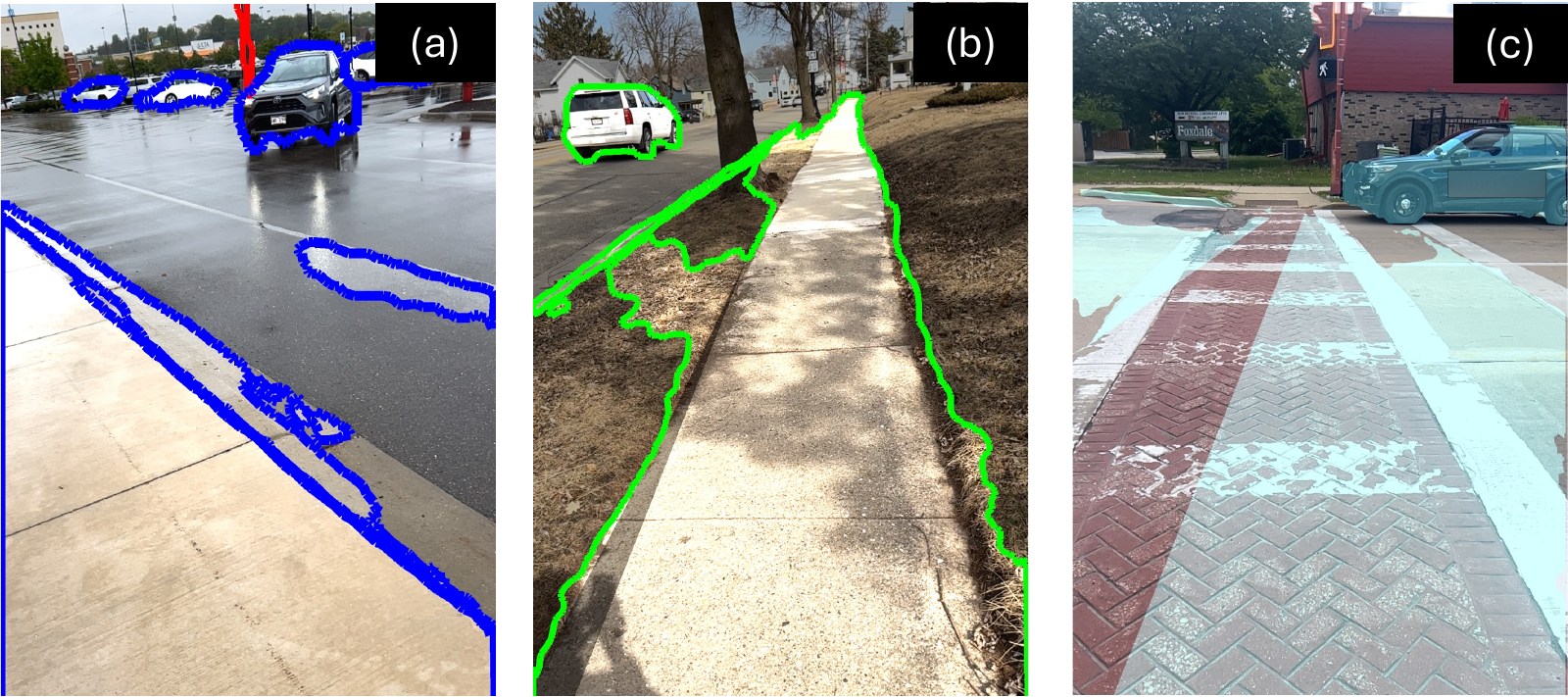}
    \caption{
    Example images showing how environmental factors affected the recognition accuracy of walkable paths.  
    (a) \textbf{Weather condition:} during rainy weather, \system{} misrecognized the reflections of water puddles on the road (augmented with blue contours) as part of the sidewalk, while failing to recognize the actual curb at the lower right as it darkened after being wet.
    (b) \textbf{Lighting and shadows:} under strong sunlight filtered through trees, \system{} incorrectly recognized bright patches on the grass as part of the sidewalk (augmented with green contours).
    (c) \textbf{Nonstandard road markings and textures:} a red-brick crosswalk with white stripes was only partially recognized, and the adjacent area within two white lines was mistakenly recognized as part of the crosswalk (augmented with cyan overlays).
    }
    \Description{
This figure (with three subfigures) lists out three example images for the three factors influencing the accuracy of NavSight's object recognition, respectively:
(a) Weather condition: during rainy weather, NavSight misrecognized the reflections of water puddles on the road (augmented with blue contours) as part of the sidewalk, while failing to recognize the actual curb at the lower right as it darkened after being wet.
(b) Lighting and shadows: under strong sunlight filtered through trees, NavSight incorrectly recognized bright patches on the grass as part of the sidewalk (augmented with green contours).
(c) Nonstandard road markings and textures: a red-brick crosswalk with white stripes was only partially recognized, and the adjacent area within two white lines was mistakenly recognized as part of the crosswalk (augmented with cyan overlays).
    }
    \label{fig:misrecognition}
\end{figure*}

Participants attributed recognition errors on walkable paths to three environmental factors: 
(1) \textbf{Weather Conditions}:
Three participants (P5, P8, P11) observed that the recognition accuracy of walkable paths decreased during rainy weather, as wet surfaces changed color and puddles changed the path appearance. Consequently, \system{} misrecognized the boundaries between wet and dry surfaces or the puddles as walkable paths (Figure~\ref{fig:misrecognition}a). 
(2) \textbf{Lighting and Shadows}:
When walking in sunny weather under trees, four participants (P1, P5, P6, P12) observed that \system{} misrecognized sidewalks due to tree shadows, which created scattered dark and bright spots on the ground. \system{} often augmented bright areas near the sidewalks as part of the walkable path (Figure~\ref{fig:misrecognition}b). P5 also found that shadowed curbs were not recognized, rating \system{} somewhat accurate (3) when this occurred, compared with very accurate (4) or extremely accurate (5) when the curbs were not shadowed.
(3) \textbf{Nonstandard Road Markings and Textures}:
Three participants (P1, P11, P12) also related recognition errors to crosswalk design and surface markings. While crosswalks in the training data have white markings (parallel lines or zebra stripes) on concrete or asphalt, participants encountered crosswalks with nonstandard marking color or surface texture. For example, P12 reported that a red-brick crosswalk was only partially recognized (Figure~\ref{fig:misrecognition}c), and P11 found that a yellow-striped crosswalk was not detected at all.

\subsubsection{Attitudes Toward Errors by Consequence}
\label{subsubsec:error_attitudes}
\begin{figure*}[tbp]
    \centering
    \includegraphics[width=0.4\textwidth]{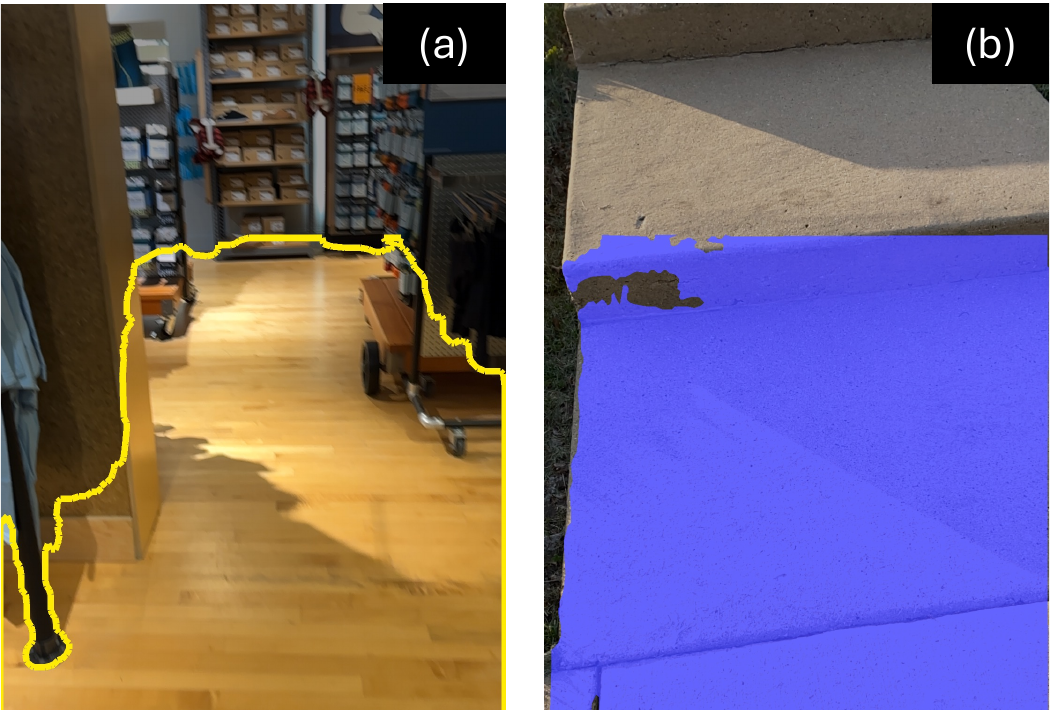}
    \caption{
    Example screenshots of recognition errors with different consequences. (a) A false positive that added information: in an indoor shopping mall, \system{} augmented the aisle between shelves as a sidewalk (yellow contour), which P1 found helpful for navigating indoors. (b) A segmentation error that removed information: the sidewalk augmentation (blue solid overlay) extended to a step, hiding the height change.
    }
    \Description{
Two screenshots from NavSight. (a) The interior of a shop, with display shelves on both sides. A yellow contour outlines the wooden floor of the aisle running between them, marking it as a sidewalk. (b) A concrete surface photographed from above. A semi-transparent blue overlay covers most of the surface, and its upper edge extends past the base of a step onto the step itself, so that the edge of the step is covered by the overlay rather than left visible.
    }
    \label{fig:asymmetric}
\end{figure*}

Participants' attitudes toward recognition errors varied by consequence. Errors that did not hide hazards were tolerated. Five participants (e.g., P4, P11) found false positives occasionally helpful because they drew attention to potential hazards. As P11 explained: ``\textit{[The augmentation] is still showing you something that could be a hazard... it makes sense [to augment it].}'' P1 even found false positives helpful in an indoor shopping mall, where \system{} augmented the aisles as sidewalks and helped her navigate between shelves (Figure~\ref{fig:asymmetric}a). She rated \system{} somewhat helpful (3) and very safe (4) that day, even though \system{} was not designed for indoor use. 
In contrast, errors that obscured hazards were concerning. P10 noted that when an inaccurate boundary overlapped a hazardous zone, it could mislead her toward it, and P9 and P12 reported that \system{} sometimes augmented a step as part of the sidewalk, hiding the height change (Figure~\ref{fig:asymmetric}b).

\subsubsection{AI Error Interpretation \& Handling}
\label{subsubsec:making_sense}
We report how low vision participants perceived and responded to AI errors, as well as the challenges they faced handling AI errors.

\textbf{Error Discovery.}
All participants noticed recognition errors by comparing the augmentations against what they saw themselves. P4 also checked augmentations against his common sense of sidewalks, noticing errors when the augmentation contained irregular scattered pieces instead of a ``straight or curved line.'' 

\textbf{Error Interpretation.}
Five participants (e.g., P6, P9) formed their own mental models of why recognition errors happened and adjusted their usage behaviors accordingly. P9 found \system{} accurate when she 
held the phone still but noticed augmentation flickering when walking. P2 and P5 attributed errors to the number of objects they selected, believing that \system{} would recognize more accurately with fewer objects selected. P6 attributed \system{}'s accuracy to her augmentation selection. After switching from solid overlay to contour enhancement, which did not block the augmented objects, she considered \system{} more accurate, noting that ``many more objects were caught on camera with those adjustments.''
These mental models led to different usage behaviors from participants (P2, P5, P9), with P9 holding the phone upright and P2 and P5 deselecting objects. 

\textbf{Reduced Trust in Inconsistent Recognition.}
Two participants (P10, P11) could not find a coherent explanation for inconsistent recognition. They found it inconsistent that \system{} augmented objects outside its supported categories (false positives) while at the same time missing selected objects (false negatives). P11 found that a drink bottle on the ground was augmented as a pole while a nearby bench was not augmented. P10 also found it inconsistent that \system{}'s false positives were random and did not apply to all instances of the same out-of-list object, such as some stair handrails being augmented as fences while others were not. This inconsistency reduced their trust in \system{}. As P10 explained: ``\textit{It lessens the faith you have [for \system{}] in the act of helping or protecting you ... and that's very important because that's what this is supposed to do.}''

\subsection{Evolving Preferences on What and How to Augment}
\label{subsec:configuration}
As participants used \system{} over time across scenarios, they developed strategies for deciding which objects to augment, how to group them, and how to augment each group. They selected objects they expected to encounter in the scene or needed for their current activity, grouped them by risk severity, and adjusted augmentations to keep them visible as the lighting changed. We elaborate below.

\subsubsection{What Objects to Augment}
\label{subsubsec:object_selection}
Participants' object selection patterns varied, with some participants tending to stick with certain sets of objects over time (e.g., P3, who kept his initial selection across all five days), while others changed object selection frequently (e.g., P4, who changed his selection on all six days). Seven participants concentrated their object set adjustments in the early days, making almost all changes in the first half of the week; for example, P5 adjusted object selection on days one, four, and five,
and kept the same set across the remaining seven days. The other five (P1, P2, P4, P10, P11) made changes evenly across their days. Despite these diverse patterns, all participants converged on a stable set of objects, and even those who kept changing made only small tweaks in the later days.

These stable sets included several common object categories. Seven participants (e.g., P2, P8) augmented \textit{common walkable paths} (i.e., crosswalks, sidewalks). Three participants (P2, P8, P12) augmented \textit{common moving objects} (i.e., vehicles, pedestrians) as they believed that they were ``definitely going to encounter'' these objects and needed to pay close attention to them (P2). P4 and P11 also consistently augmented \textit{common tripping hazards} (i.e., curbs and rail tracks) as these hazards were low to the ground and harder to detect.

Beyond this stable set, participants augmented other objects for specific environments, activities, or experiences. 
Nine participants (e.g., P7, P12) selected objects based on what was present in the scene. P10 selected what she anticipated encountering before her walk, since she could ``kind of tell what there was right there,'' and deselected fire hydrants in her rural neighborhood, where there were none. P4 added trash cans on trash collection day when ``a lot of them were out'' and deselected them on other days. Five participants (e.g., P2, P8) augmented objects relevant to their current activity. P7 and P8 selected buses when catching a bus, P6 augmented pedestrian lights when crossing streets, and P2 selected benches when looking for a place to sit and deselected it afterward. 
Two participants (P7, P11) added objects after a near-miss with objects they did not select. For example, P7 added bicycles to \textit{Augmentation Group I} after a bicycle came toward her on the bike path the previous day,
and P11 added pedestrians after nearly running into people on a bike path. 

Beyond \system{}'s supported object categories, participants also identified objects they wanted to augment. They wanted \system{} to augment surface-level changes, such as stairs (7/12; e.g., P2, P9), potholes and uneven terrain (7/12; e.g., P1, P10), and sidewalk cracks (P3, P6, P7, P9), and to expand \system{}'s coverage to rural elements, such as dirt roads (P1, P7), gravel paths (P6, P7), and tree roots (P6). When P1 walked a rural trail, where the ground was unpaved gravel and the main obstacles were rocks and trees, she found that no walkable paths or hazards were augmented, and reported: ``\textit{the app was not detecting the way it normally did so I felt very insecure using it.}''


\subsubsection{How to Group Objects}
\label{subsubsec:grouping}
In contrast to their frequent adjustments to object selection, participants' preferences on object group assignments remained consistent across the week. Ten of the twelve participants changed object group assignments on at most one day, and four of them (P1, P3, P9, P12) never did so. 
 Participants demonstrated consistent object grouping patterns. Most participants (7/12; e.g., P4, P8) grouped objects by their \textit{risk level}, assigning objects more critical to safety to one group and augmenting them using a more prominent augmentation. However, participants judged risk level differently.
Four participants (P5, P8, P9, P12) grouped by motion, separating moving objects (e.g., vehicles, bicycles) from stationary ones, because the former were ``not always as easy to see at a distance'' and posed higher risk (P5).
P4 grouped by height, assigning foot- and knee-level tripping hazards (e.g., curbs, fire hydrants) to one group and higher bumping hazards (e.g., poles) to the other, since lower hazards were harder to notice and posed greater risk.

Participants also proposed adding more augmentation groups for more granular object distinction (6/12; e.g., P2, P9). P12 suggested a third group for objects not directly related to safety but still ``good to know'' for scene awareness (e.g., fire hydrants, utility boxes). P4 and P9 further recommended a dedicated group and color for walkable areas so they could locate these areas at a glance.

In contrast, three participants (P1, P7, P10) preferred placing all important objects in the same group. 
P1 and P10 preferred augmenting all safety-critical objects equally. P10 started with two groups in the first three days, but moved every object into one group on the fourth day, only distinguishing whether an object was hazardous or not. 
P7 found determining group assignments distracting. As she explained: ``\textit{[\system{}] is really pretty complex... at first I was overwhelmed by all the choices and then as I started thinking about it, I had to think now really what is important to me and why would I use this app... I was thinking too many choices might make this less useful.}'' As a result, they recommended having a single augmentation group so they only needed to decide whether to augment each object category.


\begin{figure*}[tbp]
    \centering
    \includegraphics[width=\textwidth]{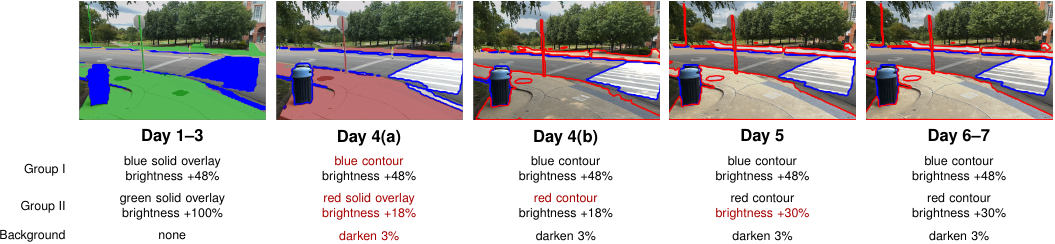}
    
  \caption{
  An illustration of P11's evolving augmentation selection over the study, shown on the same reconstructed scene. On Days~1--3, she used solid overlays for both augmentation groups (Group~I blue, Group~II green). On Day~4 she switched both groups to contour enhancement, since the solid overlays occluded the augmented objects, and darkened the background to make the augmentations stand out. She made these changes in two steps, first converting Group~I to a contour and recoloring Group~II from green to red (Day~4(a)), then converting Group~II to a contour (Day~4(b)). On Day~5 she raised Group~II's brightness from +18\% to +30\% to make it more visible, and kept this design for the rest of the study.
  }
  \Description{
This figure illustrates how participant P11 changed her augmentation design over the course of the study, rendered on a single reconstructed street scene so the design changes can be compared. Five stages are shown left to right, each labeled by day, with a table below listing the color, form, and brightness of her two augmentation groups and the background at each stage. On Days 1 to 3, both groups used solid overlays that fill the objects they mark: Group I in blue at +48\% brightness, Group II in green at +100\% brightness, with no background darkening. On Day 4, she changed both groups from solid overlays to contours that trace object outlines instead of filling them, because the solid overlays covered and hid the objects underneath, and she darkened the background by 3\% so the augmentations stood out. This happened in two steps: on Day 4(a) she converted Group I to a blue contour and recolored Group II from green to a red solid overlay at +18\% brightness; on Day 4(b) she converted Group II to a red contour. On Day 5 she raised Group II's brightness from +18\% to +30\% to make it more visible. She kept this final design, blue and red contours over a slightly darkened background, for the rest of the study.
  }
  \label{fig:design_change}
\end{figure*}

\subsubsection{How to Augment Objects}
\label{subsubsec:aug_design}
Participants changed their augmentation selections frequently throughout the study, adjusting them on two to six of their days.
Unlike object selections, for which most participants concentrated their changes in the early days (Section~\ref{subsubsec:object_selection}), participants' augmentation changes were spread across their days.
Participants learned over days which augmentations made objects easier to see, and adapted them to keep objects visible as the lighting and background changed. We summarize their augmentation selection strategies below.

\textbf{Trade-off between Visibility and Occlusion.}
Participants adjusted the augmentations to ensure visibility while minimizing occlusion. Seven participants (e.g., P1, P2) selected contour enhancement on their first day of use and five (e.g., P4, P11) selected solid overlays. Despite the visibility of solid overlays, four participants (P4, P6, P7, P11) changed to contour enhancement over the study since solid overlays occluded the augmented objects while the contours defined the object better (P7). For example, P11 started with solid overlay because the contour was too thin to see until she adjusted its thickness, whereas the solid overlay covered the whole object and she ``could see [it] right away'' without customization. After she became familiar with customizing the contour enhancement, she switched to it on her fourth day because the solid overlay blocked the object. Figure~\ref{fig:design_change} shows P11's day-to-day augmentation selection in the week.


\textbf{Adapting Augmentation to Lighting, Background, and Eye Comfort.}
Most participants (10/12; e.g., P2, P6) changed their augmentation selection based on the brightness of the scene, which varied with the weather (sunny vs. gloomy) and the time of day (morning vs. late afternoon). Six participants (e.g., P4, P11) changed augmentation colors to keep them visible against bright scenes, and four participants (P1, P2, P6, P12) darkened the background to reduce its brightness. For example, P1 darkened the background in the early afternoon, and around sunset she removed the darkening and brightened the augmented objects instead. Participants also chose colors that contrasted with the scene (P4, P12), avoiding colors that blended into the background, such as cyan against the blue sky (P4), white against clouds (P12), and red against a red fence (P12). Their color preferences differed. Under bright sunlight, P4 found yellow and green harder to see than blue and red, while P10 found yellow more visible than blue or red.

Beyond visibility, two participants (P1, P8) switched colors during longer sessions to reduce eye strain. As P8 described: ``\textit{when I was using the app for a longer period of time, switching the colors helped ... for my eyes to kind of just like refresh it and like take a break.}''

\subsection{Social Acceptability}
\label{subsec:social_acceptability}

\begin{figure*}[tbp]
    \centering
    \includegraphics[width=0.55\textwidth]{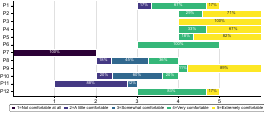}
    
  \caption{
  Distribution of each participant's daily ratings of perceived comfort using \system{} in public, from 1 (not comfortable at all) to 5 (extremely comfortable). Each bar is centered on that participant's median rating. 
  }
  \Description{
A horizontal diverging bar chart of perceived comfort using NavSight in public, with one bar per participant from P1 to P12. Each bar is centered on that participant's median rating, and colored segments show the percentage of that participant's ratings at each level, from 1 (not comfortable at all) to 5 (extremely comfortable).
P3 gave a rating of 5 on all of his days, P6 gave 4 on all of hers, and P7 gave 1 on all of hers, the lowest of any participant. P9, P5, and P2 also rated high, at 89, 82, and 71 percent of their days at 5. P11 is the second lowest, with 88 percent of her ratings at 2 and 12 percent at 3. P8 and P10 have medians of 3, and P1, P4, and P12 have medians of 4 or above. P7 is the only participant to rate NavSight not comfortable at all (1) on any day.
  }
  \label{fig:comfort}
\end{figure*}

Ten participants felt comfortable using \system{} in public (median $\geq 3$, \textit{somewhat comfortable}), praising it for reducing social conspicuousness by keeping their visual impairment less apparent to others (Section~\ref{subsubsec:concealing}). However, P7 and P11 rated it below three and raised concerns over phone use (Section~\ref{subsubsec:phone_use_perception}).


\subsubsection{Concealing Visual Impairments}
\label{subsubsec:concealing}
Eight participants (e.g., P1, P9) felt that \system{} improved social acceptability by not revealing their visual condition. Unlike a white cane, which signaled their visual impairment and created a sense of isolation (P2, P8, P9), holding a phone was viewed as a common behavior in public. As P9 said: ``\textit{if I have a cane for walking, everyone looks at you. It's very embarrassing to me.}'' She rated \system{} extremely comfortable in public (5) in eight of her nine daily surveys, feeling that no one would notice she was visually impaired.

While acknowledging \system{}'s discreetness, P4 wished that \system{} could signal his visual condition to others to avoid misunderstanding. After accidentally bumping into a pedestrian, he had to show his cane to explain, and suggested adding a signal in \system{} to indicate the user's visual condition on demand.

\subsubsection{Concerns Over Phone Use}
\label{subsubsec:phone_use_perception}
\system{} also introduced social concerns about how others might interpret participants' phone usage. Six participants (e.g., P8, P11) worried that holding the phone up could be mistaken for filming or taking photos, causing social discomfort. As P8 shared: ``\textit{I did not feel uncomfortable using the app. The only time I hesitated was if I was taking a picture of an oncoming car or bike I waited until they had passed or were too far away to get any distinctive image of the person in case, someone objected to my taking their picture with my phone.}'' P7 was even questioned by a passing pedestrian about what she was doing. She rated \system{} not comfortable at all (1) on all seven days, reporting that she did not like the appearance of videotaping strangers without their approval.

This concern depended on how many people were around. Four participants (P4, P5, P7, P8) felt less comfortable in crowded places, where more people might think they were filming. As P5 noted: ``\textit{in very public areas it can be weird to use as it may look like you are filming people.}'' 
However, P11 felt the opposite: in crowded places, she blended in among others using their phones, while in less crowded places her phone use stood out.

P7 and P11 also noted that walking with the phone held up might make others think they were distracted or inattentive. As P11 said: ``\textit{Having [the phone] in front of you can make people suspicious [of you not paying attention].}'' P7 viewed this discomfort as partly age-related: ``\textit{Older people aren't used to doing that. It's [a] taboo in the old culture to do that [use phone while navigating].}''
As a result, P7 felt discouraged from using \system{} for navigation, and P11 only used it in familiar and crowded areas to reduce the chance of misunderstanding.


\subsection{Usability Issues in the Wild}
\label{subsec:usability_issues}
The diary study revealed practical usability issues that may affect the long-term adoption of \system{}, including participants mistaking their own configuration for a malfunction, augmentations becoming hard to see under adverse weather, and \system{} consuming power and blocking other assistive applications. These issues are often not captured in controlled lab studies, yet can influence PLV's actual usage of the system in daily life.

\subsubsection{Darkened Background Mistaken for a Malfunction}
\label{subsubsec:darken_background}
While darkening the background suppressed scene brightness (Section~\ref{subsubsec:aug_design}), two participants (P2, P9) fully darkened the background so that the screen turned black when no objects were augmented, and mistook it for a malfunction. P9 left the background fully darkened for three days, reporting high distraction as she was ``not sure what happened,'' and realized it only at the end of the third day: ``\textit{I somehow didn't know that I set a dark background and changed it back.}'' P2 similarly mistook it for a malfunction and resolved it only after contacting the research team.

\subsubsection{Augmentation Visibility under Adverse Weather} 
\label{subsubsec:augmentation_visibility}
We found that weather conditions affected not only object recognition accuracy (Section~\ref{subsubsec:env_factor_acc}) but also participants' ability to perceive visual augmentations.
In sunny conditions, ten participants (e.g., P7, P11) found it difficult to see the phone display due to screen glare and limited screen brightness, which reduced augmentation visibility. As P11 noted: ``\textit{If I really was relying on [\system{}], I wouldn't feel so comfortable in a bright sunny situation because I couldn't really see it that well.}'' Three participants (P2, P7, P9) stopped using \system{} when the scene was too bright.
P1 also noted that in rainy weather, reflections from streetlights and car headlights on wet surfaces (e.g., rain puddles) created glare on the phone screen, making augmentations harder to perceive.
To address these visibility issues, P7 and P9 suggested adding audio cues to supplement visual feedback when the screen is difficult to see.

\subsubsection{Power Consumption and App Compatibility}
\label{subsubsec:power_consumption}
Consistent with our technical evaluation of \system{}'s power consumption (Section~\ref{par:system_overhead}), three participants (P2, P4, P5) noted that their phones heated up and drained battery quickly during use. P2 also disliked that \system{} could not run simultaneously with other assistive phone applications. As she relied on another mobile application to check pedestrian signals, she had to switch back and forth between the two applications when crossing streets.
\section{Discussion}
\label{sec:discussion}
In this paper, we investigated how PLV use AR visual augmentations for real-world outdoor navigation. We found that augmentations reshaped how participants perceived the outdoor scenes, simplifying them into walkable and non-walkable regions and extending their visual reach, while also dividing their attention between the phone screen and their surroundings. PLV actively configured what objects to augment and how, adapting their choices to the scene, their activities, and the lighting. Recognition failed under real-world conditions such as rain, lighting and shadows, and nonstandard road texture; some participants formed their own mental models of these errors and adapted their usage strategies.
In this section, we discuss how \system{} served as a complement to existing low-vision aids, design implications for AR systems for low vision in outdoor navigation, the challenges of deploying AI-powered assistive tools in the real world, and the limitations and future directions.

\subsection{\system{} as a Complement to Existing Low-Vision Aids}
\label{subsec:complement_tool}
\system{} was designed to complement existing low-vision aids rather than replacing them, serving a different purpose from conventional tools such as a white cane. A cane provides immediate haptic feedback of ground-level obstacles within arm's reach \cite{dos2021electronic, dos2021systematic}, whereas \system{} supported scene interpretation at greater range, enabling participants to notice objects earlier (Section~\ref{subsubsec:visual_reach}), scan for approaching traffic from different directions (Section~\ref{subsubsec:moving_objects}), and locate walkable regions (Section~\ref{subsubsec:simplify_walk}).

All six participants who carried a cane used \system{} alongside it for the additional information, and their existing aids in turn covered what \system{} missed. \system{} occasionally rendered an inaccurate sidewalk augmentation over hazardous regions (Sections~\ref{subsubsec:misrecognition} and \ref{subsubsec:error_attitudes}) and could not recognize surface hazards such as cracks and uneven tiles (Section~\ref{subsubsec:object_selection}), both of which a cane could detect. \system{} also augmented objects without conveying semantic information, such as recognizing pedestrian signals but not whether they indicated stop or go, so P2 used a different application to check the signal at every crossing (Section~\ref{subsubsec:power_consumption}). When \system{} was unusable, such as under adverse weather (Section~\ref{subsubsec:augmentation_visibility}), participants fell back on their own vision and existing aids. This pattern echoes the prior finding that mobile navigation applications for BLV people are used to supplement rather than replace traditional tools \cite{bleau2026exploring}. These usage experiences suggest that AR navigation aids should complement users' existing aids, covering the information those aids cannot provide, and taking into account the physical resources and attention they already occupy.

%
%

\subsection{Design Implications for AR Systems for Low Vision in Outdoor Navigation}
\label{subsec:design_implications}
Throughout the study, participants proposed a wide range of improvements and desired features for \system{}. We summarize and expand on these design implications for future AR systems as low-vision aids in outdoor navigation.

\subsubsection{Automatically Adapt to Context}
\label{subsubsec:automatically_adapt}
\system{} allowed users to freely assign objects to augmentation groups and choose corresponding augmentation designs. Participants used this flexibility throughout the study, selecting objects based on what was present in the scene and what they were doing, and adjusting augmentations as the scene changed (Section~\ref{subsec:configuration}). While they appreciated the agency to control their configurations, they found these frequent manual adjustments burdensome and overwhelming.
Future AR systems should automatically suggest configurations from context (e.g., users' tasks) while maintaining user control. This is especially important in outdoor environments, where objects move unpredictably and leave users no time for fine-grained customization \cite{lee2021understanding}. For example, future systems could incorporate activity-recognition models \cite{fathi2011understanding, spriggs2009temporal, pirsiavash2012detecting, chakraborty2014context, wang2020symbiotic} to identify task-relevant objects, and adapt augmentations against ambient lighting and background colors of the scenes \cite{hincapie2015smartcolor, zhang2021color}. 

\subsubsection{Provide Intelligent Support for Real-World Configuration \& Debugging} 
\label{subsubsec:intelligent_diagnose}
While \system{} let participants configure what to augment and how, participants could not easily tell what they had configured, and sometimes mistook their own settings for a system malfunction, especially in real-world settings. P2 and P9 fully darkened the background and took the black screen for a malfunction (Section~\ref{subsubsec:darken_background}). Resolving this was difficult, as P9 did not realize the cause until the third day and P2 resolved it only after contacting the research team. This echoes prior findings that, when an AI-powered assistive tool fails, users are left guessing whether the problem comes from their own settings or from the system \cite{kosa2026not}.
Future AI-powered AR systems could provide intelligent support for adjusting configurations and debugging problems in real-world use. An AI agent with access to the augmented view and the user's current configuration could answer queries such as ``\textit{why is the screen black?}'' to distinguish a problem with the user's settings from a system failure. Such an agent could also surface problems proactively \cite{kosa2026not, peng2025morae}, such as by confirming a large configuration change with the user before applying it, or by detecting an unusual augmented view and prompting the user to check their settings.


\subsubsection{Provide Options Across Platforms} 
\label{subsubsec:provide_options}
Choosing a platform for an AR navigation aid requires balancing multiple factors, such as availability, cost, social acceptability, and the benefits of the form factor.
We deployed \system{} on mobile phones because they are widely available, affordable, and socially acceptable. However, the phone form factor divided user attention from the real world and required hand-based interactions. Participants thus suggested smart glasses as the app platform, echoing prior findings that blind and low-vision users favor hands-free, gaze-aligned displays \cite{gamage2023blind}. 
Smart glasses (e.g., Ray-Ban Meta glasses) would address both problems and are increasingly used for visual assistance and low vision rehabilitation \cite{li2022scoping, waisberg2024meta, jimenez2025retiplus}, but they remain more expensive and less widely owned than phones. They are also more conspicuous and less socially acceptable \cite{hoogsteen2023holistic, gamage2023blind}, whereas holding a phone did not reveal participants' visual impairment (Section~\ref{subsubsec:concealing}). Future AR navigation aids should provide options across platforms and convey the trade-offs of each one, so that users can choose the one that fits their own priorities.

\subsection{AI in the Wild: Challenges of AI-powered Assistive Systems in the Real World}
\label{subsec:ai_in_the_wild}
We contribute one of the first real-world deployments of an AI-powered mobile AR aid for people with low vision and identify unique challenges of field deployment. Based on these challenges, we discuss four directions for future systems: expanding recognition beyond discrete objects, prioritizing the visual challenges shared by PLV and recognition models, supporting trust calibration, and matching evaluation metrics to the consequence of errors.

\subsubsection{Expand Recognition Beyond Discrete Objects}
\label{subsubsec:expand_discrete}
\system{} supported outdoor navigation by recognizing and augmenting important objects (Section~\ref{subsec:objects}). However, in real-world use, participants wanted \system{} to further convey hazards that are not discrete objects but attributes of the walking surface, such as cracks, slopes, and uneven tiles (Section~\ref{subsubsec:object_selection}). Some participants inferred these hazards from the ``holes'' in the sidewalk augmentations (Section~\ref{subsubsec:seg_cues}), but this workaround is unreliable given the unstable sidewalk augmentation (Section~\ref{subsubsec:misrecognition}). This demand for surface hazards was especially vital in rural environments, where the main hazards were unpaved ground, rocks, and transitions between surfaces.
Therefore, in addition to discrete objects, future systems should recognize the properties and irregularities of the walking surface (e.g., cracks, height changes, transitions between surfaces) by incorporating deep-learning methods that detect road defects \cite{tabernik2020segmentation, katsamenis2024deep}. Depth estimation from RGB images \cite{yang2024depth} or additional depth sensors could also be considered to detect surface-level changes such as curbs and steps \cite{yang2018unifying}.

\subsubsection{Prioritize Shared Visual Challenges for PLV and AI}
\label{subsubsec:prioritize_shared}
Outdoor navigation involves objects that are challenging for both PLV and computer vision. Participants relied on \system{} to notice objects before reaching them, but these objects appeared small in \system{}'s camera frame, which limited the range at which \system{} could recognize them (Section~\ref{subsubsec:visual_reach}). Environmental conditions also created challenges for both participants and \system{}. After rain, wet surfaces changed color and obscured the boundary between sidewalks, curbs, and roads, making it hard for participants to stay on the sidewalk (Section~\ref{subsubsec:simplify_walk}) and causing \system{} to augment the wet-dry boundaries as the path (Section~\ref{subsubsec:env_factor_acc}).
This overlap is especially concerning because \system{} was least reliable exactly when participants' own vision could not meet their needs. Future AI-powered assistive systems should prioritize these shared challenges. For example, they could construct datasets of visually challenging objects, such as objects that are small \cite{cheng2023towards} or in low light \cite{loh2019getting}, and apply image enhancement \cite{li2020all, qian2024allweather, tasnim2025normalizing} to mitigate the effect of weather and lighting before inference.


\subsubsection{Support Trust Calibration in AI Models}
\label{subsubsec:trust_calibration}
When \system{} made recognition errors, participants had no way to know why. They developed their own explanations, attributing errors to how they held the phone, how many objects they selected, or the augmentations, and acted on these mental models (Section~\ref{subsubsec:making_sense}). When participants could not form a coherent explanation, they trusted \system{} less.
These results extend prior findings that blind and low-vision people struggle to verify camera-based AI and blame themselves for its failures \cite{hong2024understanding, sakib2026explainable, alharbi2024misfitting} to real-time AR contexts. 
However, unlike the query-based applications in prior work, where users can confirm a result by re-capturing an image, \system{} augments a continuous camera stream where errors are intermittent and cannot be re-examined.
Future AI-powered AR systems should give users an accessible way to verify AI errors and calibrate their trust in real time, for example, visually conveying recognition confidence by rendering uncertain augmentations differently from confident ones \cite{tomsett2020rapid}.


\subsubsection{Match Evaluation Metrics to the Consequence of Errors}
\label{subsubsec:match_metrics}
Our fine-tuned model outperformed the baseline on mAP, mAP50, and mAP75 (Section~\ref{subsec:implementation}, Table~\ref{tab:recognition_accuracy}). \camchange{However, because mAP averages across classes, the high accuracy on objects with well-defined boundaries (e.g., cars) masked the low accuracy on safety-critical classes, such as sidewalks and crosswalks (Appendix~\ref{app:per-class}). Moreover,}
in daily use, participants judged errors by their consequences. Errors that added information were tolerable or even useful, whereas errors that obscured a real hazard were dangerous (Section~\ref{subsubsec:error_attitudes}). 
\camchange{Common metrics like mAP do not reflect the importance of each class or how users judge the consequences of different errors, as they weigh every class and every error equally \cite{padilla2021comparative}.}
These findings suggest that AI-powered assistive systems should select evaluation metrics matching the consequence of errors in the tasks. For outdoor navigation, \camchange{evaluation should prioritize the accuracy of safety-critical classes, and} prioritize minimizing errors that conceal hazards, including missed hazard instances and boundary errors that mark a hazard as walkable, over errors that augment an unimportant object.

\subsection{Limitations and Future Directions}
\label{subsec:limitations}
Our work has several limitations.
First, we only prototyped \system{} on iOS and conducted the diary study with iPhone users. While the study allowed us to gather insights on PLV's real-world outdoor experience of mobile AR, it may omit challenges related to phone models or operating systems.
Second, due to the difficulty of conducting long-term field deployment with PLV, we only included 12 participants. While this is comparable to prior field deployments with blind and low-vision users \cite{gonzalez2024investigating, zhao2018face}, it may not capture the full range of PLV's experiences, needs, and preferences. 
Third, our helpfulness, safety, and accuracy measures were self-reported and relied on participants' impressions, which could be affected by their memory or other subjective factors.
Fourth, our one-week deployment was not long term and could not reveal patterns that may emerge over months, such as whether participants would abandon \system{} because of real-world usability issues.
Finally, we were unable to quantitatively analyze the causes of recognition inaccuracies because we did not log users' camera feeds or recognition results to preserve privacy. 
Future work could deploy \system{} across more platforms and with larger, more diverse samples of PLV over longer periods, incorporate objective task-performance measures, and collect real-world egocentric navigation data to examine the causes of recognition errors and how PLV's experiences vary across platforms, individuals, and environments. 
\section{Conclusion}
In this paper, we conducted a one-week diary study with 12 people with low vision to investigate their real-world usage of \system{}, a research prototype mobile AI-powered AR application for outdoor navigation. Participants found \system{} helpful for simplifying scenes, noticing hazards and moving objects, and extending their visual reach, while also dividing their attention between the screen and their surroundings. Participants adapted what and how to augment to each scenario, formed their own explanations for \system{}'s recognition errors, and found \system{} more socially acceptable than conventional low-vision aids but noted that it could be mistaken for filming. We also identified environmental factors that affected recognition accuracy and usability issues in daily use.
Based on these findings, we discuss the real-world challenges and distill design implications for future AR and AI-powered systems for low vision in outdoor environments.

\begin{acks}
Research reported in this publication was supported by the National
Eye Institute of the National Institutes of Health under Award
Number~R01EY037100. The content is solely the responsibility of
the authors and does not necessarily represent the official views of
the National Institutes of Health.
\end{acks}

\bibliographystyle{ACM-Reference-Format}
\bibliography{main}

\newpage
\appendix
\onecolumn

\section{Per-class Accuracy}
\label{app:per-class}

\begin{table}[h]

\caption{Per-class instance-segmentation accuracy on the test set, comparing the fine-tuned \textit{YOLO11l-seg-outdoor} model with the baseline \textit{YOLO11l-seg} model (pre-trained on MS-COCO without fine-tuning), grouped by the four navigation challenges (Section~\ref{subsec:objects}). \textit{Instances} is the number of ground-truth instances per class in the test set. \textit{FNR} (false negative rate) is the proportion of ground-truth instances the model misses, calculated as $\mathit{FN}/(\mathit{TP}+\mathit{FN})$. \textit{FDR} (false discovery rate) is the proportion of the model's predictions that are false positives, calculated as $\mathit{FP}/(\mathit{TP}+\mathit{FP})$. We report FNR and FDR for the fine-tuned model under the deployment thresholds (confidence = 0.4, IoU = 0.5).
The fine-tuned model outperforms the baseline on all classes except bench. A baseline score of 0.000 indicates a class absent from MS-COCO that the baseline cannot recognize (e.g., curb, sidewalk, crosswalk). $^{*}$MS-COCO has a single traffic light class and does not distinguish between pedestrian light and vehicle traffic light, so the baseline produces the same score for both.}
\label{tab:per-class-accuracy}
\begin{tabular}{@{}l r rr rrr rrr@{}}
\toprule
\multirow{2}{*}{\textbf{Challenge / Class}} & \multirow{2}{*}{\textbf{Instances}} & \multicolumn{5}{c}{\textbf{Fine-tuned}} & \multicolumn{3}{c}{\textbf{Baseline}} \\
\cmidrule(lr){3-7} \cmidrule(lr){8-10}
 & & \textbf{FNR} & \textbf{FDR} & \textbf{AP50} & \textbf{AP75} & \textbf{AP} & \textbf{AP50} & \textbf{AP75} & \textbf{AP} \\
\midrule
\multicolumn{10}{@{}l}{\textbf{Staying on walkable paths}} \\
\quad Curb                     &  2766 & 0.465 & 0.236 & \textbf{0.603} & \textbf{0.268} & \textbf{0.297} & 0.000 & 0.000 & 0.000 \\
\quad Sidewalk                 &  2615 & 0.523 & 0.311 & \textbf{0.503} & \textbf{0.172} & \textbf{0.221} & 0.000 & 0.000 & 0.000 \\
\midrule
\multicolumn{10}{@{}l}{\textbf{Avoiding obstacles}} \\
\quad Rail track               &    89 & 0.674 & 0.383 & \textbf{0.333} & \textbf{0.110} & \textbf{0.154} & 0.000 & 0.000 & 0.000 \\
\quad Fence                    &  1857 & 0.763 & 0.337 & \textbf{0.327} & \textbf{0.102} & \textbf{0.138} & 0.000 & 0.000 & 0.000 \\
\quad Bench                    &    47 & 0.681 & 0.464 & 0.296 & 0.055 & 0.119 & \textbf{0.386} & \textbf{0.148} & \textbf{0.193} \\
\quad Sewer drain              &    99 & 0.697 & 0.400 & \textbf{0.374} & \textbf{0.178} & \textbf{0.209} & 0.000 & 0.000 & 0.000 \\
\quad Fire hydrant             &    16 & 0.250 & 0.200 & \textbf{0.852} & \textbf{0.778} & \textbf{0.612} & 0.704 & 0.651 & 0.503 \\
\quad Junction box             &   177 & 0.667 & 0.372 & \textbf{0.421} & \textbf{0.155} & \textbf{0.190} & 0.000 & 0.000 & 0.000 \\
\quad Pole                     &  4579 & 0.408 & 0.225 & \textbf{0.678} & \textbf{0.238} & \textbf{0.311} & 0.000 & 0.000 & 0.000 \\
\quad Traffic cone             &    41 & 0.146 & 0.386 & \textbf{0.830} & \textbf{0.742} & \textbf{0.605} & 0.000 & 0.000 & 0.000 \\
\quad Trash can                &   202 & 0.559 & 0.350 & \textbf{0.502} & \textbf{0.255} & \textbf{0.265} & 0.000 & 0.000 & 0.000 \\
\midrule
\multicolumn{10}{@{}l}{\textbf{Crossing streets}} \\
\quad Crosswalk                &   277 & 0.791 & 0.284 & \textbf{0.354} & \textbf{0.139} & \textbf{0.167} & 0.000 & 0.000 & 0.000 \\
\quad Pedestrian light$^{*}$   &   124 & 0.524 & 0.337 & \textbf{0.566} & \textbf{0.208} & \textbf{0.279} & 0.438 & 0.202 & 0.228 \\
\quad Vehicle traffic light$^{*}$ &   552 & 0.208 & 0.289 & \textbf{0.789} & \textbf{0.393} & \textbf{0.417} & 0.438 & 0.202 & 0.228 \\
\quad Traffic sign             &   908 & 0.301 & 0.296 & \textbf{0.752} & \textbf{0.533} & \textbf{0.481} & 0.091 & 0.061 & 0.056 \\
\midrule
\multicolumn{10}{@{}l}{\textbf{Avoiding collisions with moving objects}} \\
\quad Pedestrian               &  1315 & 0.251 & 0.243 & \textbf{0.765} & \textbf{0.367} & \textbf{0.399} & 0.640 & 0.281 & 0.313 \\
\quad Bicycle                  &   195 & 0.487 & 0.286 & \textbf{0.565} & \textbf{0.131} & \textbf{0.222} & 0.499 & 0.124 & 0.195 \\
\quad Bus                      &   255 & 0.384 & 0.238 & \textbf{0.690} & \textbf{0.446} & \textbf{0.422} & 0.614 & 0.325 & 0.355 \\
\quad Car                      &  6022 & 0.170 & 0.163 & \textbf{0.869} & \textbf{0.542} & \textbf{0.519} & 0.687 & 0.342 & 0.362 \\
\quad Motorcycle               &   183 & 0.404 & 0.273 & \textbf{0.651} & \textbf{0.189} & \textbf{0.265} & 0.524 & 0.119 & 0.198 \\
\quad Truck                    &   416 & 0.462 & 0.275 & \textbf{0.620} & \textbf{0.409} & \textbf{0.393} & 0.352 & 0.247 & 0.232 \\
\midrule
\textbf{All classes}           & \textbf{22735} & --- & --- & \textbf{0.588} & \textbf{0.305} & \textbf{0.318} & 0.256 & 0.129 & 0.136 \\
\bottomrule
\end{tabular}

\Description{A table comparing per-class instance-segmentation accuracy of a fine-tuned model and a baseline model across 21 object classes, grouped into four navigation challenges. For each class it gives the number of test instances, the fine-tuned model's false negative rate and false discovery rate, and, for both models, average precision at IoU 50, IoU 75, and IoU 50 to 95. Ten navigation-specific classes such as curb, sidewalk, and crosswalk have a baseline score of zero because they do not exist in MS-COCO. The fine-tuned model scores higher than the baseline on every class except bench, which has very few instances. Pedestrian light and vehicle traffic light share the same baseline score because MS-COCO has only one traffic light class.}

\end{table}


\end{document}